\documentclass[
    12pt, 
    twoside,
    colorlinks=true,
    linkcolor=blue,
    urlcolor=blue,
    citecolor=gray
]{article}	

\usepackage[lmargin=1in,rmargin=1in,tmargin=1in,bmargin=1in]{geometry}

\usepackage{
	amsmath,			
  	amssymb,			
  	amsthm,			    
    array,              
    blkarray,           
    bm,                 
  	bookmark,        	
    caption,
	chngcntr,			
	enumerate,		    
	fancyhdr,			
	float,				
    graphicx,           
  	mathrsfs,           
  	mathtools,       	
    minted,             
	multicol,			
	multirow,			
	nicefrac,			
	relsize,			
    setspace,
	stmaryrd,			
    scalerel,
    setspace,
    subcaption,
    tabularx,
    tcolorbox,          
  	thmtools,           
	thm-restate,		
    tikz,               
	totcount,			
    todonotes,
    wrapfig,
}

\usepackage[linesnumbered]{algorithm2e}
\usepackage[super]{nth}
\usepackage[dvipsnames,svgnames]{xcolor} 
\usepackage[final]{showlabels}      

\usepackage[final,nopatch=footnote]{microtype}

\usepackage{hyperref}

\usetikzlibrary{bayesnet, calc} 
\usepackage[
	hyperref = true,      	
  	backend  = biber,      
  	sorting  = none,       	
  	style  = numeric     
]{biblatex}
\SetKw{Initialize}{Initialize}

\setminted[python]{linenos=true, frame=lines, highlightcolor=lancasterLightYellow!20}

\definecolor{lancasterGrey}{RGB}{190,192,194}
\definecolor{lancasterBodyGrey}{RGB}{85,86,86}
\definecolor{lancasterRed}{RGB}{181,18,27}
\definecolor{lancasterBlack}{RGB}{35,31,32}
\definecolor{lancasterLightGrey}{RGB}{233,236,237}

\definecolor{lancasterDarkGreen}{RGB}{85,120,105}
\definecolor{lancasterLightGreen}{RGB}{134,153,120}
\definecolor{lancasterBeige}{RGB}{186,182,162}
\definecolor{lancasterDarkGrey}{RGB}{50,65,71}
\definecolor{lancasterLightBlue}{RGB}{127,170,190}
\definecolor{lancasterLightPurple}{RGB}{100,96,108}
\definecolor{lancasterLightRed}{RGB}{194,103,99}
\definecolor{lancasterLightYellow}{RGB}{227,203,139}
\definecolor{lancasterLightOrange}{RGB}{225,171,108}

\definecolor{UnivColor}{RGB}{212,69,0}

\newcommand{\gemlib}{\texttt{gemlib}}

\newcommand{\code}[1]{\mintinline{python}{#1}}
\newcommand{\Px}{\mathbb{P}}

\newcommand{\state}[1]{\mathrm{#1}}
\newcommand{\tx}[2]{[\mathrm{#1}\mathrm{#2}]}
\newcommand{\sidx}[1]{\scaleto{\state{#1}}{0.75ex}}
\newcommand{\stx}[2]{\scaleto{\state{#1}\state{#2}}{0.75ex}}

\newlength{\overwritelength}
\newlength{\minimumoverwritelength}
\newcommand{\overwrite}[3][red]{%
  \settowidth{\overwritelength}{$#2$}%
  \ifdim\overwritelength<\minimumoverwritelength%
    \setlength{\overwritelength}{\minimumoverwritelength}\fi%
  \stackrel
    {%
      \begin{minipage}{\overwritelength}%
        \color{#1}\centering\small #3\\%
        \rule{1pt}{9pt}%
      \end{minipage}}
    {\colorbox{#1!10}{\color{black}$\displaystyle#2$}}}

    \newtcolorbox{highlightbox}[1]{colback=red!5!white, colframe=red!75!black,fonttitle=\bfseries, title={#1}}

\newcolumntype{L}{>{$}l<{$}} 
\newcolumntype{C}{>{$}c<{$}} 

\title{\gemlib: probabilistic programming for epidemic models}
\author{Alin Morariu, Jessica Bridgen, Chris Jewell}

\begin{document}

\maketitle

\section*{Abstract}
\gemlib~is a Python library for defining, simulating, and calibrating Markov state-transition models. Stochastic models are often computationally intensive, making them impractical to use in pandemic response efforts despite their favourable interpretations compared to their deterministic counterparts. \gemlib~decomposes state-transition models into three key ingredients which succinctly encapsulate the model and are sufficient for executing the subsequent computational routines. Simulation is performed using implementations of Gillespie's algorithm for continuous-time models and a generic Tau-leaping algorithm for discrete time models. \gemlib~models integrate seamlessly with Markov Chain Monte Carlo samplers as they provide a target distribution for the inference algorithm. Algorithms are implemented using the machine learning computational frameworks JAX and TensorFlow Probability, thus taking advantage of modern hardware to accelerate computation. This abstracts away computational concerns from modellers, allowing them to focus on developing and testing different model structures or assumptions. The \gemlib~library enables users to rapidly implement and calibrate stochastic epidemic models with the flexibility and robustness required to support decision during an emerging outbreak. 

\section{Introduction}\label{sec:intro}

Mathematical models of infectious disease spread through populations have become a key component of managing outbreaks at national and international levels. They are typically used for estimating critical outbreak metrics such as the reproduction number, growth rates, and the efficacy of interventions \cite{Sofonea2022Epidemic}. Models may also be used to project forwards in time, providing an opportunity to identify \emph{in-silico} the likely effect of control strategies, such as social distancing and vaccination, as load on healthcare systems \cite{alahmadi2020}.

One of the advantages of infectious disease modelling is the ability to draw together many sources of outbreak data, such as case incidence and prevalence, human mobility, demographics and socioeconomics, and pathogen genetics \cite{shadboltEtAll2022}. However, the complexity of models required to do this comes at a high price: models are complex to implement efficiently on modern computing hardware, and require a high level of skill in mathematics, statistics, and computer science in addition to epidemiology \cite{deanglisEtal2015}. In the absence of suitable high-level toolkits, implementing models from scratch is a slow, error-prone process that detracts from the main tasks of building models, challenging them with data, and making improvements and innovations where necessary \cite{swallow2022pandemicmodelling}.

Software toolkits for epidemic modelling should allow for the specification of multiple models. This is essential in order to fully explore the range of epidemic possibilities where ensemble simulations are standard practice for scenario modelling \cite{Leng2021SocialBubbles, Moore2021Vaccination}. They encompass a wide range of model-specific inputs (e.g. parameter values, population structures, environmental conditions) but stop short of including a diversity of models themselves. Although a single model may be adapted to address different questions, doing so often requires substantial modification to account for variation. Adopting a Bayesian approach across temporal, spatial, and behavioural factors associated with a single outbreak. Policymakers benefit most when results are informed by sets of independent models with complementary methodologies and diverse data streams \cite{Sofonea2022Epidemic}, since this reduces dependence on any one set of assumptions and highlights the role of both stated and implicit hypotheses in shaping outputs. To support such ensemble approaches, software should not confine modellers to rigid, pre-built mathematical structures; rather, it should empower them to encode diverse structures and assumptions without restricting access to inference or calibration tools.

The current state of epidemic modelling reveals a substantial gap between the limited, “out-of-the-box” inference methods available for easy application and the continually emerging, state-of-the-art algorithms developed within the statistics community. This gap has resulted in application-specific packages and specialist tools with limited reusability, which contradict modern data science best practices \cite{Pruim2023Fostering}. Our goal is to address this challenge by developing a framework that enables seamless interaction between modellers and statisticians. Within this framework, novel methodology can be incorporated as modules, with interoperability ensured by a predictable, mathematically principled model interface. In turn, inference algorithms can be applied across the broad array of model classes supported in the library, allowing modellers to experiment with them while retaining responsibility for appropriate usage. By generalizing model representation, the framework reduces reliance on narrowly focused packages or software that restrict the scope of usable models. Instead, we establish a minimal set of requirements for model interaction, thereby providing the statistics community with a clear specification for implementing new algorithms within the library. To do this we take inspiration from \code{nimble} \cite{deValpine03042017nimble} and focus on separating out "one time" set-up steps (e.g. defining the model structure) from repeated "run time" steps such as simulating a stochastic system one step forward in the process. 

As data streams grow in size and complexity, there is a growing tension between ease of implementation and computational performance. High-level, expressive programming languages such as Python and R are preferred due to their readability and maintainability. Meanwhile, low-level languages such as C++ demand significantly greater implementation effort and customization specific to a problem. The practicality of epidemic models is naturally tied to the computational performance \cite{thompsonEtAl2020} that is achieved by the framework they are implemented in. We addresses this challenge by building our framework on top of highly performant back-end libraries such as JAX and TensorFlow Probability, thereby eliminating the need for low-level implementation while maintaining computational efficiency. This design choice further enables distributed computation and seamless use of high-performance computing resources, allowing large-scale models to be operationalized. Moreover, models developed within the framework can be integrated directly into modern data science pipelines and leverage contemporary computing hardware like Graphical Processing Units (GPUs) or Tensor Processing Units (TPUs), ensuring both accessibility and scalability.

This paper introduces the Python library, \gemlib; (see \href{https://gitlab.com/gem-epidemics/gemlib}{Gitlab}), which is primarily designed for the rapid construction and analysis of stochastic compartmental epidemic models - the predominant class of models used in epidemiology \cite{keeling2011modeling, Sofonea2022Epidemic}. \gemlib~provides a unified framework for defining the data generating process, with a focus on flexibility, modularity, and crucially for a seamless integration into modern Bayesian workflows. A key feature of \gemlib~is the generic compartmental model specification system, which allows users to define models using reusable building blocks such as compartments, transitions, and population structures. This design philosophy prioritizes the model itself and explicitly separates the compartmental model from any algorithms that researchers may want to use for simulation or parameter inference. The result is a library that facilitates iterative refinement of data-informed epidemic models, while maintaining code clarity and reproducibility.

\subsection{Adopting a Bayesian approach}

\begin{figure}[h!]
    \centering
    \begin{subfigure}[b]{0.48\textwidth}
        \centering
        \begin{tikzpicture}
            \node[latent] (mu) {$\mu$};
            \node[latent, right=of mu] (sigma) {$\sigma$};
            \node[obs, below=of $(mu)!0.5!(sigma)$] (x) {x};
            \edge {mu, sigma} {x};
        \end{tikzpicture}
        \caption{Bayesian DAG for a normally distributed random variable $x$ with unknown mean $\mu$ and standard deviation $\sigma$}
        \label{fig:normal-dag}
    \end{subfigure}
    \hfill
    \begin{subfigure}[b]{0.48\textwidth}
        \centering
        \begin{tikzpicture}
            \node[latent] (beta) {$\beta$};
            \node[latent, right=of beta] (gamma) {$\gamma$};
            \node[latent, below=of $(beta)!0.5!(gamma)$] (SIR) {SIR};
            \node[obs, below=of SIR] (Data) {Cases};
            
            \edge {beta, gamma} {SIR};
            \edge {SIR} {Data};
        \end{tikzpicture}
        \caption{Bayesian DAG for the SIR compartmental model described in \cite{kerMcK1927}}
        \label{fig:sir-dag}
    \end{subfigure}
    \label{fig:dags}
    \caption{Examples of Bayesian models represented via directed acyclic graphs. Nodes represent random variables in the model while edges show the conditional dependence between nodes.}
\end{figure}
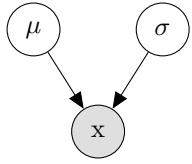
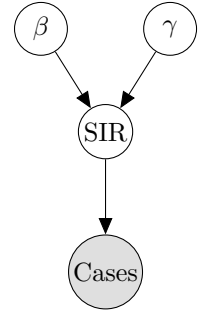

A Bayesian approach offers a number of distinct advantages for performing inference on epidemic models \cite{Sofonea2022Epidemic, oniellEtAl1999}. A useful way to represent Bayesian models is through \textit{directed acyclic graphs} (DAGs), where nodes denote random variables and arrows encode dependencies. For example, assume that $x$ are observations from a Normal distribution with unknown mean $\mu$ and standard deviation $\sigma$. The DAG for this model is shown in Figure \ref{fig:normal-dag}. We bundle the model parameters together into a parameter set $\theta = \{ \mu, \sigma \}$. Given values for the parameters, we can simulate values for the process $x$. Additionally, we can also perform the inversion where by given observations $x$ we update the posterior distribution (as outlined by Bayes' Theorem in Equation \ref{eq:bayes-thm}). The updated posterior provides refined estimates, based on data, of parameters $\theta$ (via Bayesian inference methods such as Markov Chain Monte Carlo, Sequential Monte Carlo or otherwise). This is shown in Equation \ref{eq:bayes-thm}. 

\begin{equation}\label{eq:bayes-thm}
    p(\theta | x) \propto p(x|\theta) p(\theta) =  \mathcal{L}(\theta ; x) p(\theta)
\end{equation}

This is a simple observational model where model parameters map directly to data and therefore constructs the data generating model. However, epidemic models have a more intricate structure with competing sub-processes which make performing the inversion more difficult. Consider the canonical SIR model first described by \cite{kerMcK1927} where individuals begin as susceptible to infection, then become infected, followed by removed (either by recovery or death). In this process, it is impossible to observe each transition event precisely and even a reported case only provides partial information about the process as a whole. A DAG representation gives us two benefits. Firstly, it outlines the data generating model by making an assumption about the rate at which infections and removals happen. Secondly, allows for inference to be performed since it includes probabilistic links between the latent status of individuals (i.e. infection and recovery times), model parameters, and observed variables (such as case counts). We can represent this posterior symbolically via a conditional probability decomposition as such: 
\begin{equation}
    p(\theta, \text{latent states} | \text{observed cases}) \propto p( \text{observed cases} |\text{latent states} , \theta) p(\text{latent states} |\theta) p(\theta)
\end{equation}
From this, we can see an essential distinction: unlike observational models, epidemic processes rely on parsimonious parametrizations to explain non-trivial system behaviour \cite{Sofonea2022Epidemic} that are associated with latent stochastic process. Unfortunately, the combination of these complex processes means that epidemic models are difficult to embed within a DAG directly as shown in Figure \ref{fig:sir-dag}. While specific models \textit{can} be encoded to do this, the reusability remains limited since modifying an element of the model would involve modifying the source code and managing trickle down effects. This is precisely the problem highlighted earlier and one we aim to address in \gemlib\; by generalizing compartmental models. Effective statistical software should therefore enable users to define arbitrarily complex compartmental models and automatically compute the corresponding likelihoods, allowing them to focus on model design rather than bespoke implementations of the probabilistic elements. Modellers can then take advantage of state of the art algorithms \cite{fearnhead2024scalablemontecarlobayesian}.

\subsection{Probabilistic programming}

Probabilistic programming provides a compelling foundation for such a framework by embedding Bayesian statistics within expressive programming languages. A probabilistic program expresses a model in terms of its data generating model, and specifies a conditional inference problem by conditioning on nodes within the DAG that have been observed. The resulting program is typically declarative in nature, focusing on declaring and defining probability distributions, and the way in which they interact according to the DAG. As such, a probabilistic programming language (PPL) focuses on the model structure, abstracting away the details of the algorithms that sample from the associated probability space. Thus the analyst is freed from the laborious process of repeated writing of algorithmic code, and instead is allowed to focus on the statistical analysis at hand. This approach has been highly successful in general Bayesian statistics: probabilistic programming languages such as WinBUGS, STAN, PyMC, and Nimble have become standard for rapid development and fitting of, in particular, Bayesian hierarchical models \cite{Lunn2000WinBUGS, Lee2017stan, Vincent-pymc3, deValpine03042017nimble}.

Probabilistic programming therefore has many advantages to offer for the infectious disease modeller, especially since the algorithms for fitting models to incomplete observations of an outbreak are complex to programme. Literature published before and after Covid19 has a consensus that such a framework is urgently needed to enable rapid response to disease outbreaks, the development of trustworthy and reproducible model implementations, and lower the barrier to entry for students approaching infectious disease modelling for the first time \cite{deanglisEtal2015, thompsonEtAl2020, swallow2022pandemicmodelling,shadboltEtAll2022, mitchell2022fairdata, Sofonea2022Epidemic}. To address these concerns, we propose that a successful PPL for infectious disease modelling must have the following features:
\begin{enumerate}
    \item a transparent software pattern for \textit{describing} an infectious disease model, which may be implemented as continuous- or discrete-time stochastic, or deterministic process;
    \item \emph{automatically generated} prior simulation and probability density functions bespoke to the model description, abstracting algorithm logic away from the analyst;
    \item the ability to \emph{run models on any hardware}, with accelerators such as GPUs automatically used with no further requirement for software modification;
    \item a \emph{loosely-coupled} software architecture, which provides interfaces and patterns for developers to easily extend and build on the library. 
\end{enumerate}

\begin{figure}
    \centering
    \begin{tikzpicture}
            
        \coordinate (A) at (-4,0);
        \coordinate (B) at ( 4,0);
        \coordinate (C) at (0,4);
        \newcommand{\levels}{JAX,TensorFlow Probability, \gemlib}
        
        \draw (A) -- (C);
        \draw (B) -- (C);
        
        \newcounter{levels}
        \foreach \A in \levels {\stepcounter{levels}}
        
        \foreach[count=\y from 0] \A in \levels {
            \draw ($(A)!\y/\the\value{levels}!(C)$) 
                -- ($(B)!\y/\the\value{levels}!(C)$) 
                node[midway,above] {\A};
        }
    \end{tikzpicture}   
    \caption{The \gemlib~library was built on top of JAX and TensorFlow Probability as both back end libraries provide useful functionality for computational tasks specific for epidemic modelling. JAX lays the foundation for array-oriented numerical computation while TensorFlow Probability provides access to probabilistic constructs which can be used in combination with other probabilistic programming langauges.}
    \label{fig:gemlib-structure}
\end{figure}
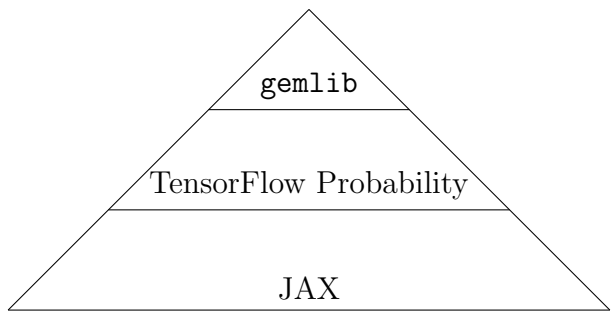

In \gemlib~we provide a mathematically coherent framework, from which the software architecture is derived. This allows for arbitrarily complex Markov state-transition models, with a suite of 3 integrators for continuous- and discrete-time stochastic processes, as well as a deterministic ODE solver. These three integrators have a common interface, with the advantage that the analyst can not only easily explore models with different numbers of compartments and transition rate specifications, but also switch in and out integration schemes with minimal changes to the code. Moreover, for the stochastic models, an automatically-generated probability density/mass function is provided, allowing principled statistical inference using the analyst's algorithm of choice. Furthermore, the library is constructed on top of powerful machine learning libraries like JAX and TensorFlow that help ensure scalability and computational efficiency (see. Figure \ref{fig:gemlib-structure}).

\section{Markov state transition models}\label{sec:markov_models}

\begin{figure}
    \centering
    \begin{subfigure}[b]{0.4\textwidth}
        \centering
        \begin{tikzpicture}
            \filldraw[fill=green!20!white, draw=black] (0,0) rectangle node{\large S} (1,1) ;
            
            \draw[thick,->] (1, 0.5)  -- node[above] {$\lambda^{\stx{S}{I}}$} (2.5, 0.5) ;
            
            \filldraw[fill=lancasterRed!20!white, draw=black] (2.5,0) rectangle node{\large I}(3.5,1);
    
            \draw[thick,->] (3.5, 0.5) -- node[above] {$\lambda^{\stx{I}{R}}$} (5, 0.5) ;
            
            \filldraw[fill=lancasterLightBlue!20!white, draw=black] (5,0) rectangle node{\large R} (6,1);
        \end{tikzpicture}
        \caption{SIR model \cite{kerMcK1927}}
        \label{fig:STM-sir}
    \end{subfigure}
    
    \begin{subfigure}[b]{0.4\textwidth}
        \centering
        \begin{tikzpicture}
            \filldraw[fill=green!20!white, draw=black] (0,0) rectangle node{\large S} (1,1) ;
            
            \draw[thick,->] (1, 0.5) -- node[above] {$\lambda^{\stx{S}{A}}$}  (2.5, 1.5) ;
            \draw[thick,->] (1, 0.5) -- node[below] {$\lambda^{\stx{S}{I}}$}  (2.5, -0.5) ;
            
            \filldraw[fill=lancasterLightRed!20!white, draw=black] (2.5,1) rectangle node{\large A} (3.5,2);
            \filldraw[fill=lancasterLightRed!40!white, draw=black] (2.5,-1) rectangle node{\large I} (3.5,0);

            \draw[thick,->] (3.5, 1.5) -- node[above] {$\lambda^{\stx{A}{R}}$} (5, 0.55) ;
            \draw[thick,->] (3.5, -0.5) -- node[below] {$\lambda^{\stx{I}{R}}$} (5, 0.45) ;

            \filldraw[fill=lancasterLightBlue!20!white, draw=black] (5,0) rectangle node{\large R} (6,1);
            
        \end{tikzpicture}
        \caption{SAIR, branched model with asymptomatic (A) and symptomatic (I) infection \cite{Liu2021}}
        \label{fig:STM-SAIR}
    \end{subfigure}
    \hfill
    \begin{subfigure}[b]{0.4\textwidth}
        \centering
        \begin{tikzpicture}
            \filldraw[fill=green!20!white, draw=black] (0,0) rectangle node{\large S} (1,1) ;
            
            \draw[thick,->] (1, 0.5) -- node[above] {$\lambda^{\stx{S}{I}}$}  (2.5, 0.5) ;
            \draw[thick,->] (1, 0.5) -- node[below left] {$\lambda^{\stx{S}{V}}$}  (2., -0.75) ;
            
            \filldraw[fill=lancasterRed!20!white, draw=black] (2.5,0) rectangle node{\large I}(3.5,1);
            \filldraw[fill=lancasterLightPurple!40!white, draw=black] (1.5,-1.75) rectangle node{\large V}(2.5,-0.75);
    
            \draw[thick,->] (3.5, 0.5) -- node[above] {$\lambda^{\stx{I}{R}}$} (5, 0.5) ; 
            \draw[thick,->] (2.25, -0.75) -- node[below right] {$\lambda^{\stx{V}{R}}$}  (3, 0) ; 
            
            \filldraw[fill=lancasterLightBlue!20!white, draw=black] (5,0) rectangle node{\large R} (6,1);
            
            \draw[thick,->, bend left=-45] (5.5, 1) to node[below] {$\lambda^{\stx{R}{S}}$} (0.5, 1);
        \end{tikzpicture}
        \caption{SVIRS, branched and cyclic model with vaccination \cite{AlexanderEtAl2004}}
        \label{fig:STM-SVIRSS}
    \end{subfigure}

    \caption{Variants of Markov transition models commonly seen in epidemic modelling. The SIR model described in \cite{kerMcK1927} is often seen as the canonical epidemic model. Modifications to the model are used to more closely reflect real-life observed epidemic processes which can include branched models with vaccinations, asymptomatic infectious, and waning immunity to allow for re-infections. While these graphs were used for modelling human diseases, they are also applicable to veterinary ones.}
    \label{fig:STMS}
\end{figure}
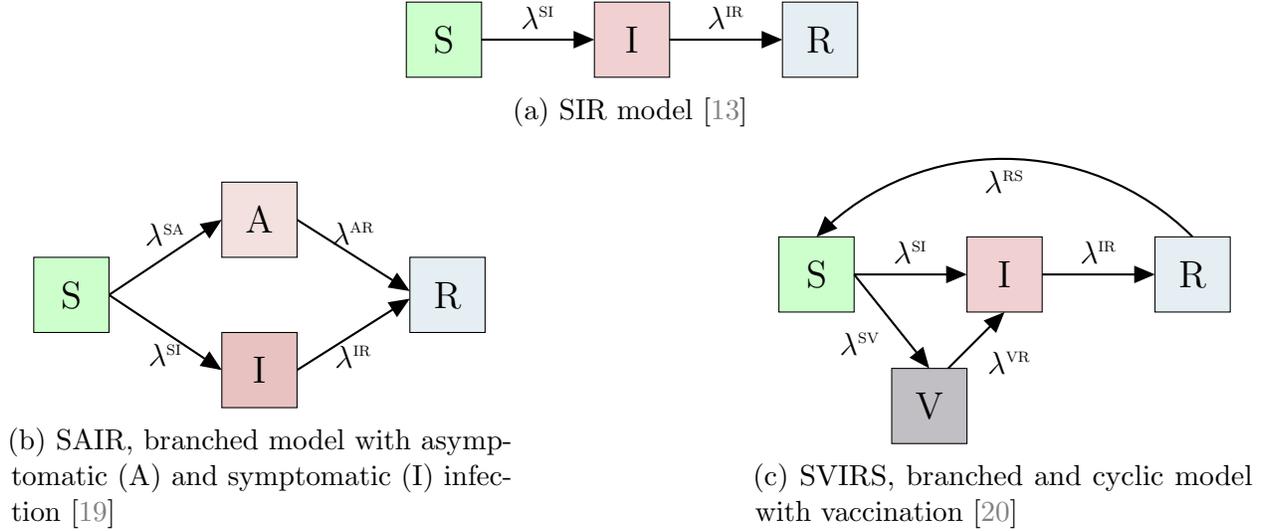

In this section, we briefly review infectious disease models in terms of a more general class of Markov state-transition model. Since differential equation solvers are already included in many PPLs, our focus is on stochastic models (though \gemlib~includes an ODE solver for completeness). Our aim is to describe these models in terms of common epidemiological terms, using an intuitive model building workflow from which the software design results.

Epidemiologists will typically begin building an infectious disease model by specifying a set of epidemiological compartments which represent different stages of the infection process undergone by an individual in the population. Models may range from as few as two compartments representing susceptible and infected \cite{RimEtAl23}, to highly elaborate models with many compartments required for vector-borne disease dynamics \cite{Porco99}. We denote the set of epidemiological compartments as $\mathcal{X}$.

Furthermore, it is common to divide the population into a number of \emph{strata} $i=1,\dots,M$, such as age- or sex- groups, or spatial regions. This is true even of individual-level models where each stratum is composed of exactly one individual. We denote by $x^{\sidx{q}}_{it}$ the number of individuals in compartment $q$ and stratum $i$ at time $t$. For convenience, we denote the population state $X_t$ at time $t$ as a $M \times |\mathcal{X}|$ matrix.
\begin{equation}\label{eq:state_t}
X_t = \left(
\begin{array}{ccc}
x^{\sidx{a}}_{1t} & \cdots & x^{\sidx{z}}_{1t} \\
\vdots & \ddots & \vdots \\
x^{\sidx{a}}_{Mt} & \cdots & x^{\sidx{z}}_{Mt} \\
\end{array}
\right)
\end{equation}

Transitions between epidemiological compartments are now specified, each associated with a transition rate. We denote the set of transitions $\mathcal{Z}$, and for compartments $q, r \in \mathcal{X}$ we denote the transition rate $\lambda^{\stx{q}{r}}(t, X_t)$ as a function of time $t$ and the epidemiological state of the population $X_t$ at time $t$. The collection of transition rate functions is then denoted by $\Lambda (t, X_t)$. Graphical depictions of three epidemic models are shown in Figure \ref{fig:STMS}, illustrating the potential for linear, branched, and cyclical topologies.

The graphical representations of compartments and transitions provide the outline for describing disease dynamics at a point in time, but we have not yet described how the population state evolves in time. For stochastic models, the state is evolved by recursively applying a Markov kernel
\begin{equation}
    k : (t, X_t) \mapsto ((s, X_s), Z_t)
\end{equation}
which transforms the time-state pairing $(t, X_t)$ and outputs the next time-state pairing $(s, X_s)$ \emph{and} an event matrix $Z_t$, a ${|M|\times |\mathcal{Z}|}$ matrix with elements $z^{\stx{q}{r}}_{it}$ giving the number of individuals in stratum $i$ undergoing the $\tx{q}{r}$ transition in the interval $[t,s)$.

The specification of $k$ now depends on whether the model is cast as a continuous- or discrete-time process. 

\subsection{Continuous-time}
In continuous time, we consider the infectious disease model to evolve as a continuous-time Markov Jump Process. To simulate trajectories from the model, we make use of the well-known Gillespie algorithm \cite{gillespie1977}. 

The kernel $k$ comprises a pair of Exponential and Discrete random variables. Given the population state $X_t$ at time $t$, the time to the next epidemiological event $\delta t$ is drawn
\begin{equation}\label{eq:gillespie-s-step}
\delta t \sim \mbox{Exponential}(\Vert  \Lambda (t, X_t) \Vert)
\end{equation}
where $\Vert  \Lambda (t, X_t) \Vert$ denotes the sum over all transition rates and population strata. Then, a \emph{single} transition event (i.e. a single individual in a single stratum undergoing a single transition) is chosen with probability equal to its own transition rate as a proportion of the total event rate
\begin{equation}\label{eq:gillespie-z-step}
Z_t \sim \mbox{Discrete}\left( \frac{ \Lambda (t, X_t)}{\Vert \Lambda (t, X_t) \Vert} \right)
\end{equation}
with $Z_t$ denoting a $M \times |\mathcal{Z}|$ one-hot tensor with the one entry corresponding to the chosen event, and zero entries elsewhere. The population state at time $s = t + \delta t$ is then given by
\begin{equation}\label{eq:state-update}
X_{s} = X_t + Z_t \cdot B
\end{equation}
where $B$ is the $|\mathcal{X}| \times |\mathcal{Z}|$ \emph{incidence matrix} such that each column giving the change in population state for each transition. For example, the incidence matrix for the SIR example in Figure \ref{fig:STM-sir} is
\begin{equation}\label{eq:incidence-matrix}
B = \left( 
\begin{array}{rr}
-1 &  0 \\
 1 & -1 \\
 0 &  1
\end{array}
\right).
\end{equation}

For this kernel, both $\delta t$ and $Z_t$ are conditionally independent given $t$ and $X_t$, such that the joint probability of drawing $(\delta t, Z_t)$ is
\begin{equation}
    \Px(\delta t, Z_t | t, X_t) = \Vert \Lambda (t, X_t) \odot  Z_t \Vert \exp \left\{- \Vert \Lambda (t, X_t) \Vert \delta t \right\}
\end{equation}
noting that $Z_t$ is a \emph{one-hot} tensor, which undergoes an element-wise product with $ \Lambda (t, X_t)$, effectively selecting out the element of $\Lambda (t, X_t)$ corresponding to the event which occurred.

\subsection{Discrete-time}
In discrete-time, we approximate the continuous time process by discretising time into intervals of fixed length $\delta t$. In this case, the Markov kernel $k$ proceeds by constructing a $|\mathcal{X}| \times |\mathcal{X}|$ right stochastic transition matrix for each population stratum $i$ with elements
\begin{equation}\label{eq:chain-multinomial-transition-matrix}
    \bm{p}^{\sidx{qr}}_{it} = 
    \begin{cases}
        1 - e^{\lambda^{\sidx{qr}}_i(t, X_t)\delta t} & \mbox{if } q \ne r, \; \tx{q}{r} \in \mathcal{Z} \\
        1-\sum_{r: r \ne q} \left( 1 - e^{\lambda^{\sidx{qr}}_i(t, X_t)\delta t}\right) & \mbox{if } q = r, \; \tx{q}{r} \in \mathcal{Z} \\ 
        0 & \mbox{otherwise} \\
    \end{cases}
\end{equation}
such that the rows of the resulting $|\mathcal{X}| \times |\mathcal{X}|$ tensor $P_{it}$ sum to 1.

The event tensor $Z_{it}$, analogous to that in Equation \ref{eq:gillespie-z-step}, is then obtained by sampling from a \emph{row-wise} Multinomial distribution such that for the $q$th row
\begin{equation}\label{eq:chain-multinomial-draw}
\bm{z}^{\sidx{q}}_{it} \sim \mbox{Multinomial}(\bm{x}^{\sidx{q}}_{it}, \bm{p}^{\sidx{q}}_{it}),
\end{equation}
and new state is obtained by taking the column sums of $Z_{it}$, i.e.
\begin{equation}\label{eq:chain-multinomial-state-update}
\bm{x}_{it} = \bm{1}^T \cdot Z_{it}.
\end{equation}
This process is thus replicated for each population stratum $i = 1,\dots,M$ to give the population state tensor at $s = t + \delta t$, $X_{s}$.

Straightforwardly, the probability of obtaining event matrix $Z_t$ given the state $X_t$ is then
\begin{equation}\label{eq:chain-multinomial-pmf}
  Pr(Z_t | t, X_t) = \prod_{i,q,r} p^{\sidx{qr}}_{it}.
\end{equation}

The discrete-time approximation comes about by treating all individuals as conditionally independent given the state $X_t$ at the beginning of the time-step, and allowing each individual to undergo at most 1 transition in each time-step. For practical purposes, and with a careful choice of $\delta t$ to minimise the chance of an individual undergoing more than one transition per time-step, this approximation often works well giving a computational saving where multiple events may be simulated without updating the population state (and therefore having to recalculate the transition rate matrices).

\subsection{Deterministic models}
Deterministic transmission models describe population-level dynamics of infectious diseases through systems of difference equations or ordinary differential equations (ODEs) \cite{kerMcK1927, keeling2011modeling}. This approach does not account for stochasticity in the system. In other words, given identical initial conditions, the modelled epidemic trajectory remains the same across repeated simulations.

Referring back to the SIR model previously mentioned, we can define the accompanying system of differential equations as follows: 
\begin{align}
    \frac{\textnormal{d}S}{\textnormal{d}t} &= - \frac{\beta S I}{N},\\
    \frac{\textnormal{d}I}{\textnormal{d}t} &= \frac{\beta S I}{N} - \gamma I,\\
    \frac{\textnormal{d}R}{\textnormal{d}t} &= \gamma I.
\end{align}
The simple model structure can be easily extended to account for heterogeneity in social mixing or mobility, and to capture more complicated disease dynamics by ways of extending the number of compartments and transitions in the model. Conveniently, it can also be more compactly represented using the incidence matrix $B$ and ordered transition rate functions $\Lambda$ as: 
\begin{equation}
    \frac{\textnormal{d}X}{\textnormal{d}t} = B\Lambda 
\end{equation}

Due to nonlinearity, these equations cannot be solved analytically and require numerical approximation. Standard numerical integration schemes such as the Runge-Kutta, Dormand-Prince method or Euler's method are employed for this purpose. Parameter estimation can subsequently be performed using approaches such as maximum likelihood estimation. We have included a differential equation solver within \gemlib~to provide analysts with the completeness and flexibility afforded by this modelling framework.

\section{Probabilistic inspired software design}

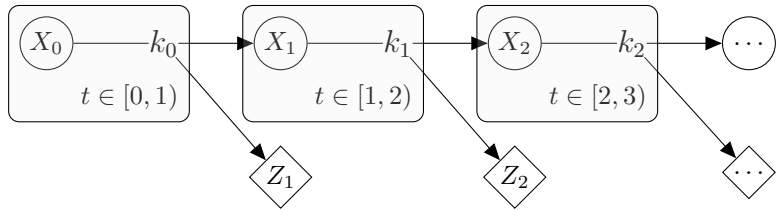
\begin{figure}[h!]
    \centering
    \begin{tikzpicture}
        \node[latent] (X0) {$X_0$};
        \node[const, right =of X0] (k0) {$k_0$};
        
        \node[latent, right =of k0] (X1) {$X_1$};
        \node[const, right =of X1] (k1) {$k_1$};
        
        \node[latent, right =of k1] (X2) {$X_2$};
        \node[const, right =of X2] (k2) {$k_2$};
    
        \node[latent, right =of k2] (Xt) {$\ldots$};
    
        \node[det, below =of X1] (z1) {$Z_1$};
        \node[det, below =of X2] (z2) {$Z_2$};
        \node[det, below =of Xt] (zt) {$\ldots$};
    
        \edge[-] {X0} {k0};
        \edge[-] {X1} {k1};
        \edge[-] {X2} {k2};
    
        \edge {k0} {X1, z1};
        \edge {k1} {X2, z2};
        \edge {k2} {Xt, zt};

        \plate [fill=gray!15, rounded corners, fill opacity=0.2] {step1} {(X0)(k0)} {$t \in [0,1)$ };
        \plate [fill=gray!15, rounded corners, fill opacity=0.2] {step2} {(X1)(k1)} {$t \in [1,2)$};
        \plate [fill=gray!15, rounded corners, fill opacity=0.2] {step1} {(X2)(k2)} {$t \in [2,3)$ };
    \end{tikzpicture}
        
    \caption{Markov transition model as propagated forward by the time-heterogenous evolution of transition kernels. The state $X_t$ is propagated forward by a time-state dependent kernel $k$. The kernel outputs the next state as well as the state change $Z_t$. Each kernel corresponds to the left-closed, right open time interval.}
    \label{fig:markov-transitions}
\end{figure}

Within \gemlib, our goal is to maintain a transparent model implementation. In practice, this means structuring the software so that it mirrors the mathematical formulation of the Markov transition kernel underlying the epidemic model. Compartmental infectious disease models are particularly well suited to this approach, as the overall epidemic process can be expressed as a sequence of Markov transition kernels. Each transition in $\mathcal{Z}$ has a state-dependent transition kernel associated with it, which can be expressed symbolically as
\begin{equation}
    k:(t, X_t) \rightarrow \left((t, X_{t}), X_{t}\right).
\end{equation}
A transition maps a time–state pair onto a new time–state pair, which naturally leads to composition between two kernels, as illustrated in \ref{eq:kernel-chaining}:
\begin{align}\label{eq:kernel-chaining}
    k \circ k &= (t, X_t) \rightarrow \left((t, X_t), Z_t \right) \; \circ \;  (t, X_t) \rightarrow \left((t, X_t), Z_t \right) \\
    &= (t, X_t) \rightarrow \left((t, X_t), [Z_t, Z_t]\right) 
\end{align}
The result is itself a kernel, together with a trace of the state changes $Z_t$ that occurred.
We formalize this kernel composition using the Chapman–Kolmogorov property. A probability kernel is defined as a mapping
\begin{align}
    k: (t,X_t) \times \mathcal{X} &\rightarrow [0,1] \label{eq:kernel-prob-mapping} \\
    ((t,X_t), \mathcal{X}) &\mapsto \Px((t,X_t) \in \mathcal{X}), \label{eq:kernel-state-mapping}
\end{align}
which can be interpreted in two ways. Equation \ref{eq:kernel-prob-mapping} computes the probability of an observed state transition, while Equation \ref{eq:kernel-state-mapping} describes the transition matrix from one time–state to another. The Chapman–Kolmogorov equation ensures that the two-step transition defined by kernel composition (Equation \ref{eq:kernel-chaining}) is itself a valid kernel \cite{lawvere1962kernels}.

The central implementation challenge lies in this time–state dependence of the probability kernel: at each step, the kernel must be reconstructed according to the current epidemiological state. The dependence means that infectious disease models are piecewise constant, time-homogeneous stochastic processes. However, we note that the model is already specified with respect to the compartments, transition rates, and population structure. As such, we want to use those components and delegate the construction of these kernels to the computer, thereby automating the creation of the epidemic process. By induction, we can show that the entire process is a Markov transition kernel so we can compute the probabilities of observing a sequence of transition events \emph{or} simulate an entire process (equivalent to traversing the DAG in both directions). The Markov transition kernel is precisely the structure that enables us to represent an epidemic model in a DAG and thus integrate compartmental models into a Bayesian workflow as we aimed to do. We can say that the process follows a distribution parametrized by an initial state and a model. 
\begin{equation*}
    \left( (t,x), z \right) \sim k \left( (t,x), z \mid \mbox{model}, t_0, x_0 \right)
\end{equation*}
In other words, the stochastic process $k$ produces a random variable which has its evolutionary dynamics governed by the infectious disease model. 

\section{Model abstraction and user interface}\label{sec:epi_ingredients}

As outlined above, the Markov transition kernel depends only on a set of initial conditions and the epidemic model. Epidemic models in \gemlib~are defined as a triplet of an integrator, transition rate functions, and a representation of the graphical model (i.e. see Figure \ref{fig:STMS}). We encode this triplet in \gemlib~using a combination of: 
\begin{itemize}
    \item Python \code{Callable}s for the transition rate functions
    \item JAX \code{Array}s for the initial state and incidence matrix
    \item TensorFlow Probability \code{Distribution} to build the Markov kernel
\end{itemize}
The result is a layered library (as in Figure \ref{fig:gemlib-structure}) where JAX handles the numerics and core operations while TensorFlow Probability provides the probabilistic constructs. 

The graphical structure of the state transition model (e.g. Figure \ref{fig:STMS}) encoded using the corresponding \textit{incidence matrix}, denoted by \( B \) as in Equation \ref{eq:incidence-matrix} (this is sometimes referred to as the stoichiometry matrix). The columns of the matrix represent the state changes associated with each transition event while the rows correspond to the epidemiological states. A detailed discussion of the construction of the incidence matrix can be found in \textcite{Black2019Importance} and \textcite{vankampen2001stoichmat}. 

The collection of transition rate functions $\Lambda$ are specified as a  \code{List} of \code{Callable}s. This closely matches the mathematical representation of an ordered set and we denote it with $[\Lambda]$. The list data structure allows for the evaluation of each function independently, and therefore can be evaluated in parallel under a multiple instruction, same data computational paradigm. 

Lastly, the time integrator, denoted by $W$,  is used to outline \textit{how} the kernel propagates the system forward and thus we define these at the \code{class} level \footnote{Each class therefore contains time-domain specific algorithms}. Stochastic models in \gemlib~are constructed using either the \code{ContinuousTimeStateTransitionModel} or the \code{DiscreteTimeStateTransitionModel} classes, with the deterministic ODE setup implemented by the function \code{ode_model}. This distinction is important as the data generating model for each time integrator is different and the implementation of the accompanying algorithms allow for additional computational efficiencies to be built into the respective classes. Consequently, each class outputs a data structure specific to its time integrator. Since continuous time models assume one event occurs per time step, returning a fully materialized state could be very memory intensive  (e.g. for individual-level models with large population sizes). 

In \gemlib, the state transition model is simply defined by these three components $G = \{W, \mathbf{B}, [\Lambda]\}$, which are sufficient for constructing the associated Markov kernel. They outline the epidemic model structure and dynamics while also providing a compact, ergonomic representation of the model. Furthermore, by adopting this representation, both the continuous- and discrete-time model classes have the \emph{same} signature. This makes it easy to swap between the two stochastic time-domains. The remainder of this section is dedicated to the specific classes in the library. 

\subsection{\code{ContinuousTimeStateTransitionModel}}
The \code{ContinuousTimeStateTransitionModel} class provides a structured interface for specifying, simulating, and evaluating probabilities for continuous-time state-transition Markov processes. The class is instantiated with an epidemic model $G$, where each component corresponds to a named argument, along with the desired number of Markov jumps for the process (specified via \code{num_steps}). Once instantiated, the model’s primary user-facing methods include \code{sample(seed)}, which generates simulated realizations of the process using the Gillespie algorithm for continuous-time Markov simulation (based on a fixed random number generating seed to ensure reproducibility), and \code{log_prob()}, which computes the log-likelihood of observed event sequences under the specified model dynamics.

\begin{minted}[linenos, highlightlines= 1]{python}
ContinuousTimeStateTransitionModel(
        transition_rate_fn,
        incidence_matrix,
        initial_state,
        num_steps,
        **kwargs)
\end{minted}

Simulation output is represented as an \code{EventList} with attributes \code{time}, \code{transition}, \code{unit} giving the time, transition index, and unit (stratum) of each successive transition event. As continuous-time models restrict the system to a single event per time step, recording events in this sparse format is considerably more efficient than storing the full event tensor $Z_{t}$ at each step. This efficiency is especially important in large-scale or individual-level models, where each event typically alters only a single entry of the state vector. The full state trajectory can be reconstructed using the \code{compute_state} method, which iteratively updates the initial state tensor (see Algorithm \ref{alg:cts-gillespie} and full reconstruction details are provided in the Supplementary Material). This approach allows users to efficiently store epidemic trajectories while maintaining seamless compatibility with downstream analyses. 

Additional introspection is available through property accessors such as \code{transition_rate_fn}, \code{incidence_matrix}, \code{initial_state}, and \code{num_steps}, which return the corresponding component. Internally, the class inherits from Tensorflow Probability's \code{Distribution} class which enables the use of vectorization and automatic differentiation that is vital for probabilistic modelling and inference. 

\subsection{\code{DiscreteTimeStateTransitionModel}}
The \code{DiscreteTimeStateTransitionModel} class provides a structured interface for specifying, simulating, and evaluating probabilities for discrete-time state-transition Markov processes. The class is instantiated with an epidemic model $G$ as before with the \textit{same} named arguments. Additional parameters such as the initial time step, \code{initial_step}, and the step duration, \code{time_delta}, allow users granular control over the discretization of time. Once instantiated, the main user interface is similar to before. Methods include \code{sample(seed)}, which generates simulated trajectories of the process by drawing from the corresponding discrete-time transition probability matrix, and \code{log_prob()} which computes the log-probability of observed transition sequences under the model’s dynamics.

\begin{minted}[linenos, highlightlines= 1]{python}
DiscreteTimeStateTransitionModel(
        transition_rate_fn,
        incidence_matrix,
        initial_state,
        num_steps,
        **kwargs)
\end{minted}

Simulation output is represented as a dense tensor of transition counts rather than an \code{EventList}. Because transitions in discrete time can occur in parallel across units, it is more efficient to store all events within a single tensor than to represent them sequentially, as in continuous-time formulations. The complete state trajectory can be reconstructed using the \code{compute_state} method, which iteratively applies the transition updates defined by the incidence matrix to the initial state tensor see Algorithm \ref{alg:chain-multinomial} and full reconstruction details are provided in the Supplementary Material). 

Users can access model components via property accessors such as \code{transition_rate_fn}, \code{incidence_matrix}, \code{initial_state}, \code{time_delta}, and \code{num_steps}. The class also provides a \code{transition_prob_matrix} method, which compute the time-dependent Markov transition probability matrix, either for the initial state or for each step in the simulated trajectory. The \code{DiscreteTimeStateTransitionModel} also inherits from Tensorflow Probability's \code{Distribution} class which enables similar workflows as before. 

In summary, each implementation encapsulates the construction of the kernel by closing over the model specification $G$. Internally, they provide implementations of the \code{sample} and \code{log_prob} methods, tailored to the underlying time regime. The incidence matrix, transition rate functions, and initial state define a compact and expressive representation for state transition models. Reducing the complexity of model formulation to these three core components, models can be efficiently constructed and modified to explore various hypotheses and scenarios. 

\subsection{\code{deterministic.ode_model}}
The \code{deterministic.ode_model} class provides a structured interface for specifying and solving differential equation systems. The specified system is solved using one of TensorFlow Probability's built-in differential equation solvers (solvers include Runge-Kutta, Dormand-Prince, or BDF \cite{keeling2011modeling}). The class is instantiated with an epidemic model $G$ as before with the \textit{same} named arguments. Additional parameters such as the initial time step, \code{initial_step}, and the step duration, \code{time_delta}, allow users granular control over the system. Since this is a deterministic system, the subsequent methods and outputs different from the model's stochastic counterparts and thus no longer require methods for sampling and  evaluating probabilities. Instead, we directly output the solution to the system as time-state pairings which can be used for subsequent analysis. 
\begin{minted}[linenos, highlightlines= 1]{python}
def ode_model(
        transition_rate_fn,
        incidence_matrix,
        initial_state,
        num_steps,
        **kwargs)
\end{minted}

\subsection{Constructing a model in \gemlib}

We demonstrate the use of \gemlib~by constructing a discrete-time stochastic metapopulation model with a static connectivity network. Such models are often used to capture the spatial dynamics of a disease outbreak. For example, livestock outbreaks modelled at the farm level or human disease outbreaks modelled at the city or regional level, whereby animal movements or commuter patterns are proxies for connectivity between spatial units. 

Here, a stochastic SIR process is used to model transmission across 3 metapopulations (units), classifying individuals into three disease states. Individuals progress from susceptible, to infected, and finally to removed according to state-dependent transition rates. For each unit $i$ at time $t$ the infection rate ($S \rightarrow I$) is proportional to the number of infected individuals in unit $i$ and the sum of the infected individuals in all other units, weighted by a connectivity matrix. The infection rate for unit $i$ at time $t$ is as follows:

\begin{equation}\label{eq:disc-si}
\lambda^{\mathrm{SI}}_i(t, x_t)  = \frac{\beta_1  x_{it}^{I} + \beta_2 (\vec{C} \cdot \vec{x}_t^{I})}{\vec{n}}
\end{equation}

where $C$ is a zero-diagonal connectivity matrix between units, $\vec{n}$ is a vector of population sizes, and $\theta = \{\beta_1, \beta_2\}$ are parameters to be estimated.

We assume the transition rate from infected to removed ($I \rightarrow R$) is known, and constant across time and population. Denoted by:

\begin{equation}\label{eq:disc-ir}
\lambda^{\mathrm{IR}}_i(t, {x}_t) = \gamma\,\mbox{day}^{-1}
\end{equation}

where $\gamma = 1/10$.

To implement this model in \code{gemlib}, we must first define the SIR state transition graph topology as a $3 \times 2$ incidence matrix. 
\begin{minted}[highlightlines = 2-4]{python}        
                              #SI  IR
incidence_matrix = np.array([[-1,  0],  #S
                              [ 1, -1],  #I
                              [ 0,  1]]) #R
\end{minted}
Next, we define our \code{initial_state} and our transition rates according to equations \ref{eq:disc-si} and \ref{eq:disc-ir}. 
\begin{minted}[linenos,highlightlines={2-9, 14}]{python}
# Define the initial state for the 3 metapopulations
initial_state = np.array(
    [  # S   I  R
        [99, 1, 0],
        [100, 0, 0],
        [100, 0, 0],
    ]
)

# Define the transition rate functions
def si_rate(_, state):
    return (
        beta1 * state[:, 1] + (beta2 * jnp.matvec(contact_network, state[:, 1]))
    ) / state.sum(axis=-1)
    
def ir_rate(_0, _1):
    return gamma
\end{minted}
We now instantiate the model using \code{DiscreteTimeStateTransitionModel} to propagate the system forward in discrete time-steps. The \code{sample} method is then used to simulate from the epidemic model. To demonstrate the stochasticity of the system we call the sample method 100 times and plot the resulting epidemic curves.

\begin{minted}[linenos]{python}
# Instantiate the epidemic model
discrete_sir = DiscreteTimeStateTransitionModel(
    transition_rate_fn=[si_rate, ir_rate],
    incidence_matrix=incidence_matrix,
    initial_state=initial_state,
    initial_step=initial_step,
    time_delta=time_delta,
    num_steps=num_steps,
    name="sir",
)

# Simulate the epidemic trajectory 100 times
stoch_events = jax.jit(
    lambda: discrete_sir.sample(sample_shape=100, seed=jax.random.key(0))
)()

# Compute the states for each trajectory
stoch_states = discrete_sir.compute_state(stoch_events)

plot_timeseries(stoch_states)
\end{minted}
\begin{figure}
    \centering
    \includegraphics[width=0.5\linewidth]{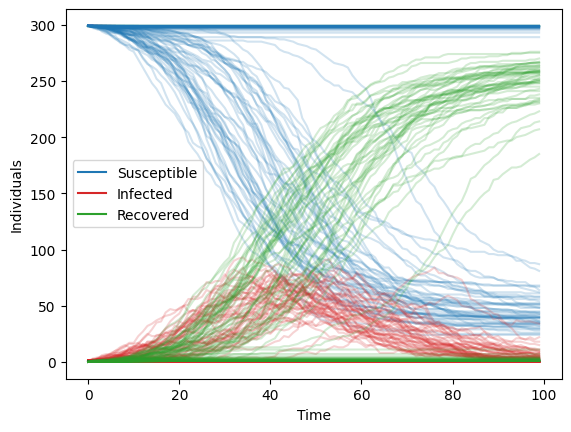}
    \caption{Epidemic timeseries for 100 trajectories, showing the number of individuals in each disease state, with values summed across the three metapopulations. }
    \label{fig:discrete=sir}
\end{figure}

We now embed our model into an instance of TensorFlow Probability's \\ \code{JointDistributionCoroutineAutoBatched}. Here, we specify the prior distributions for the parameters we wish to estimate and then instantiate the \code{DiscreteTimeStateTransitionModel}. This enables us to easily compute the log-probability of the model.

\begin{minted}[linenos]{python}
@tfd.JointDistributionCoroutineAutoBatched
def model():
    # Priors
    beta1 = yield tfd.Gamma(
        concentration=0.5,
        rate=10.0,
        name="beta1",
    )
    beta2 = yield tfd.Gamma(
        concentration=0.2,
        rate=10.0,
        name="beta2",
    )

    # Epidemic model
    gamma = 1 / 10

    def si_rate(_, state):
        return (
            beta1 * state[:, 1]
            + (beta2 * jnp.matvec(contact_network, state[:, 1]))
        ) / state.sum(axis=-1)

    def ir_rate(_0, _1):
        return gamma

    sir_model = DiscreteTimeStateTransitionModel(
        transition_rate_fn=[si_rate, ir_rate],
        incidence_matrix=incidence_matrix,
        initial_state=initial_state,
        initial_step=initial_step,
        time_delta=time_delta,
        num_steps=num_steps,
        name="sir",
    )
    yield sir_model
\end{minted}

We can now simulate an epidemic and compute the log-probability given beta1, beta2 and the simulated epidemic events.

\begin{minted}[linenos]{python}
sir_sim = model.sample(seed=jax.random.key(1))

model.log_prob(beta1=0.05, beta2=0.005, sir=sir_sim.sir)
\end{minted}

\section{\gemlib~case studies} \label{sec:examples}

In this section we showcase two examples of implementing an epidemic model and performing parameter estimation. For brevity, we omit listing any package imports. The full code is available in the Jupyter Notebooks linked at the end of the section. The purpose of these case studies is to illustrate the role \gemlib~as part of an epidemic analysis workflow and how it expedites model implementation. 

\subsection{Deterministic modelling: Covid-19 in China}
Our first case study demonstrates the implementation of a deterministic epidemic model, with noisy observations, using \code{gemlib.distributions.ode_model} by re-implementing the model proposed by \textcite{readEtAl2020} to analyse Covid19 cases in China in early 2020. The model uses a coupled set of ordinary differential equations (ODEs) to implement an SEIR metapopulation model, where metapopulations represent each of 187 cities in China. The Chinese cities are connected to each other by known data on the number of airline passengers flying between them, and each city is connected to other countries again using known airline passenger data. In the original paper, these connectivity data were subject to commercial non-disclosure and so here we make use of equivalent simulated networks. 

Below, we demonstrate the main features of \code{gemlib.distributions.ode_model} that simplify the model implementation and subsequent analysis. In particular, although \gemlib~might be considered ``Bayesian first'', here we show that it can equally be used in a frequentist paradigm following that of the original paper. The full analysis is presented in the \texttt{wuhan.ipynb} notebook available in the Supplementary Information, together with a NetCDF4 file \texttt{wuhan\_example\_data.nc} containing the data and referred to as \code{ds} in the code examples below.

We begin by defining the incidence matrix for an SEIR model, in which individuals progress from susceptible, to exposed (i.e. infected but not yet infectious), to infected, and finally to removed (i.e. recovered with solid immunity or dead). We do this as a \emph{Numpy} array, with 4 rows corresponding to the 4 (ordered) states, and columns corresponding to the $\tx{S}{E}, \tx{E}{I}$, and $\tx{I}{R}$ transitions.

\begin{minted}{python}
incidence_matrix = np.array([[-1,  0,  0],
                             [ 1, -1,  0],
                             [ 0,  1, -1],
                             [ 0,  0,  1]])
\end{minted}

The initial state of the population can now be constructed as a "batch" of 187 vectors of length 4 containing the numbers of individuals in each of the S, E, I, and R compartments in each of the Chinese cities, giving a $187 \times 4$ matrix. The model assumes that all cities begin on 1st January 2020 with their entire populations susceptible, with the exception of Wuhan which is given $I_0$ (unknown) infected susceptibles. We include a helper function \code{make_initial_state} allowing us to dynamically generate an initial state given an updated value of $I_0$ as necessary for the optimisation procedure below.

\begin{minted}{python}
WUHAN_IDX = np.where(ds["china_city"] == "Wuhan")[0][0]
initial_state = make_initial_state(
    num_initial_infectious=15.0,
    popsize=np.asarray(ds["china_population"]),
    initial_index=WUHAN_IDX,
)
\end{minted}

\textcite{readEtAl2020} now assume that the $\tx{S}{E}$ transition rate is given by
\begin{equation}\label{eq:det-se}
\vec{\lambda}^{SE}(t, x_t) = \beta \left(\vec{x}^{I}_t + K \cdot (\vec{x}^I_t \odot \vec{N}^{-1}) 
\right) \odot \vec{N}^{-1}
\end{equation}
where $K: k_{ij}$ is the (observed) number of passengers flying into city $i$ from city $j$, $\vec{N}$ is the population size of cities in China, and $\beta$ is the infection rate.

The $\tx{E}{I}$ transition rate is assumed known and common to all metapopulations
\begin{equation}\label{eq:det-ei}
\vec{\lambda}^{EI}(t, x_t) = 1/4\,\mbox{day}^{-1},
\end{equation}
and the $\tx{I}{R}$ transition rate is assumed unknown with parameter $\gamma$ again common to all metapopulations
\begin{equation}\label{eq:det-ir}
\vec{\lambda}^{IR}(t, x_t) = \gamma\,\mbox{day}^{-1}.
\end{equation}

In principle, we could now write down the set of ODEs governing our SEIR system
\begin{align*}
\frac{\mathrm{d}\vec{S}}{\mathrm{d}t} & = -x^S_t\lambda^{SE}(t, x_t) \\
\frac{\mathrm{d}\vec{E}}{\mathrm{d}t} & = x^S_t\lambda^{SE}(t, x_t) - x^E_t\lambda^{EI}(t, x_t) \\
\frac{\mathrm{d}\vec{I}}{\mathrm{d}t} & = x^E_t\lambda^{EI}(t, x_t) - x^I_t\lambda^{IR}(t, x_t) \\
\frac{\mathrm{d}\vec{R}}{\mathrm{d}t} & = x^I_t\lambda^{IR}(t, x_t).
\end{align*}
However, by using \gemlib~we can quickly construct this system without explicitly specifying the differential equations using \code{gemlib.deterministic.ode_model}. As with the stochastic model counterparts, all that is required is to supply the incidence matrix, initial state matrix, and a list of transition rate functions. Since the solution to the ODE system depends on the parameters $\beta$, $\gamma$, and $I_0$, and is essentially a mathematical function of them, we represent the model as function in code

\begin{minted}[highlightlines=22-29]{python}
def covid_ode_model(beta, gamma, I_0, num_steps=30):
    """Build a network-based ODE SEIR model"""

    initial_state = make_initial_state(
        I_0, popsize, WUHAN_IDX
    )

    def se_rate(_, state):
        within_city = state[:, 2]
        between_city = jnp.matvec(
            ds["china_connectivity"] / popsize[:, np.newaxis], state[:, 2]
        )
        return beta * (within_city_prev + between_city_flux) / ds["china_population"]

    def ei_rate(_0, _1):
        return 0.25

    def ir_rate(_0, _1):
        return gamma

    # N.B. default underlying solver is tfp.math.ode.DormandPrince
    return ode_model(
        transition_rate_fn=[se_rate, ei_rate, ir_rate],
        initial_state=initial_state,
        incidence_matrix=incidence_matrix,
        num_steps=num_steps,
        initial_time=0.0,
        time_delta=1.0,
    )
\end{minted}

To fit the model, \textcite{readEtAl2020} now assume that the observed number of new case detections on day $t$ for Chinese city $i$ is
$$
y_{it} \sim \mbox{Poisson}\left(\phi_i \left(x^{R}_{i(t+1)} - x^{R}_{it}\right)\right)
$$
where $\vec{\phi}$ is a vector of cases reporting probabilities for each Chinese city, which here we assume known for exposition purposes. Then, the number of case reports in non-Chinese country $i$ on day $t$ is assumed to be
$$
v_{it} \sim \mbox{Poisson}\left(\frac{W_{i\cdot} \cdot (\vec{\phi} \cdot \vec{x}^{I}_{t})}{\vec{N}}\right)
$$
where $W \in \mathbb{R}^{M \times N}$ is a matrix such that $w_{ij}$ is the mean number of passengers flying from Chinese city $j$ to international city $i$. 

To represent this observation process, we embed our deterministic SEIR metapopulation model within an instance of \code{JointDistributionCoroutineAutoBatched}

\begin{minted}[highlightlines=7]{python}
def wuhan_joint_model(beta, gamma, phi, I_0, num_steps = 31):

    @tfd.JointDistributionCoroutineAutoBatched
    def observation_model():

        # Solve ODEs
        seir = covid_ode_model(beta, gamma, initial_infectives, num_steps)

        # Case ascertainment vector is 1 apart from at WUHAN_IDX
        case_ascertainment_prob = (
            jnp.ones_like(ds["china_population"])
            .at[WUHAN_IDX]
            .set(phi)
        )

        # China observation model
        exp_obs_cases = expected_observed_cases(seir.states)
        yield tfd.Poisson(
            rate= exp_obs_cases * case_ascertainment_prob,
            name="china_cases",
        )

        # World observation model
        china_prev = (
            seir.states[..., 2] / ds["china_population"] * case_ascertainment_prob
        )
        world_rate = jnp.matvec(ds["world_connectivity"], china_prev)
        yield tfd.Poisson(rate=world_rate[:-1], name="world_cases")

    return observation_model
\end{minted}

To obtain the maximum likelihood estimate (MLE) $\hat{\theta} = (\hat{\beta}, \hat{\gamma}, \hat{\phi}, \hat{I_0})$ may be obtained by maximisation with a standard optimiser such as TensorFlow Probability's \code{nelder_mead_minimize}
\begin{minted}[linenos]{python}
@jax.jit
def obj_fn(data, params):
  model = covid_joint_model(*unstack(params))
  return -model.log_prob(*data)

opt = tfp.optimizer.nelder_mead_minimize(
    lambda par: obj_fn((ds["china_cases"], ds["world_cases"]), par), 
    initial_vertex=np.array([0.1, 0.1, 0.1, 0.1]),
    func_tolerance=1.0e-4,
)
\end{minted}
After asserting convergence (\code{assert opt.converged}), the parameter estimates may be obtained (\code{opt.position}).

The above code example demonstrates a further advantage of the \gemlib~library -- the ability to use JAX's ``just-in-time'' compilation feature, obtained by decorating the objective function \code{obj_fn} by \code{@jax.jit}. In our testing on a modern GPU-enabled laptop, we found this gave approximately a 4x speedup over the non-compiled code with further gains with increasing problem size (in numbers of cities or model compartments). 

\begin{figure}
    \centering
    \includegraphics[width=0.7\textwidth]{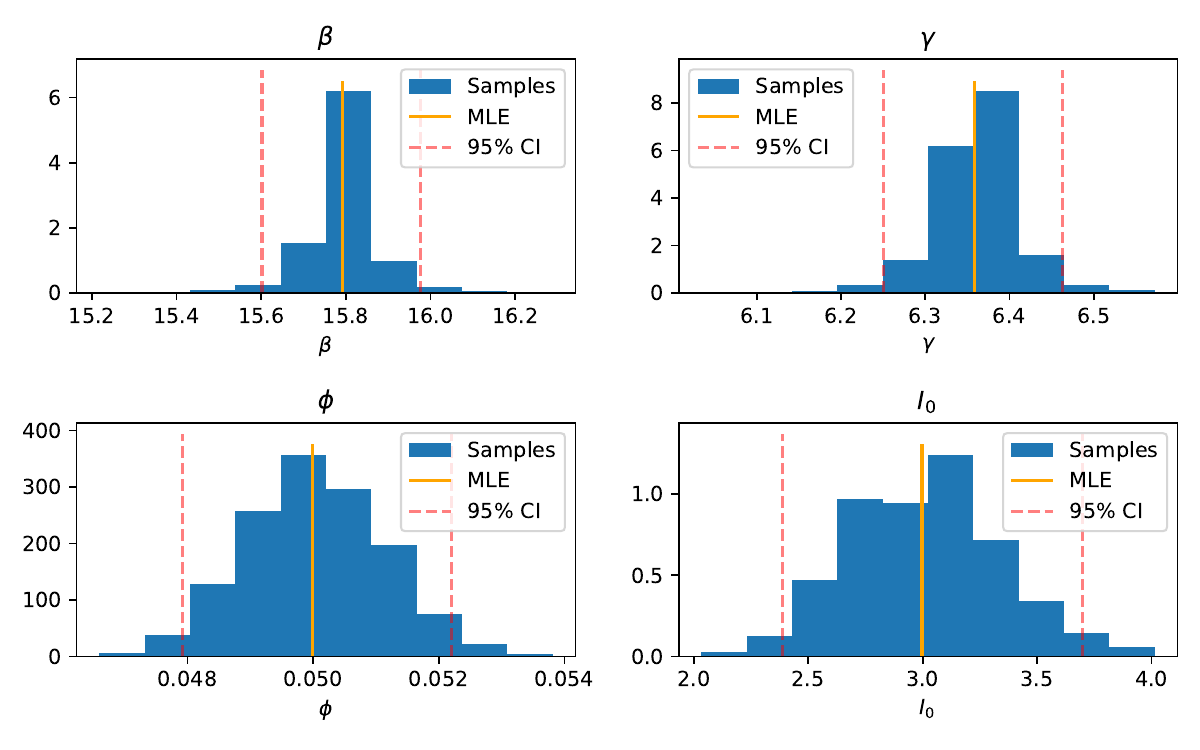}
    \caption{Estimated parameters and bootstrap distributions for the parameters of the Covid19 deterministic frequentist analysis after \textcite{readEtAl2020}.}
    \label{fig:wuhan_estimates}
\end{figure}

The compatibility with \code{JointDistributionCoroutineAutoBatched} now allows us to implement easily the bootstrap algorithm for confidence intervals as used in the original paper. First, we simple use the \code{wuhan_joint_model.sample} method to draw 1000 samples from the observation model given the MLE $\hat{\theta}$ 
\begin{minted}{python}
predictive_model = wuhan_joint_model(*opt.position)
bootstrap_sims = predictive_model.sample(
    sample_shape=1000,
    seed=jax.random.key(0),
)
\end{minted}
followed by using JAX's \code{jax.vmap} function to map the optimiser over each bootstrap sample of the observations
\begin{minted}{python}
@jax.vmap
def sample_optimize(simulated_data):
    opt = tfp.optimizer.nelder_mead_minimize(
        lambda p: obj_fn(simulated_data, p),
        initial_vertex=jnp.stack(estimated_parameters),
        func_tolerance=1.0e-3,
    )
    return opt.position
bootstrap_estimates = sample_optimize(bootstrap_sims)
\end{minted}
with the resulting sampling distributions shown as histograms in Figure \ref{fig:wuhan_estimates}. In the accompanying Jupyter notebook, we further show how a predictive distribution of the epidemic may be computed from the bootstrap parameter estimates, giving a full quantification of uncertainty in a frequentist fashion.

\subsection{Stochastic modelling: Highly pathogenic avian influenza}

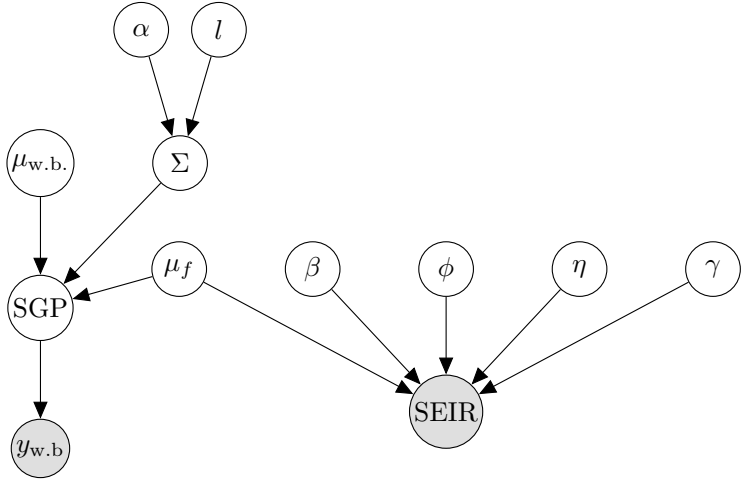
\begin{figure}[ht]
  \centering
  \resizebox{0.6\textwidth}{!}{%
    \begin{tikzpicture}
        
        
        \node[latent] (mu_farm) {$\mu_f$};
        \node[latent, right=of mu_farm]     (beta)      {$\beta$};
        \node[latent, right=of beta]        (phi)       {$\phi$};
        \node[latent, right=of phi]         (eta)       {$\eta$};
        \node[latent, right=of eta]         (gamma)     {$\gamma$};
        
        \node[obs, below=of phi] (SEIR) {SEIR};
        
        
        \edge {mu_farm, beta, phi, eta, gamma} {SEIR};
        
        
        \node[latent, left=of mu_farm, yshift=-0.5cm] (SGP) {SGP};
        \node[obs, below=of SGP] (bird-deaths) {$y_{\text{w.b}}$};
        
        \node[latent, above=of SGP]          (mu-wb)   {$\mu_{\text{w.b.}}$};
        \node[latent, right=of mu-wb] (cov) {$\Sigma$};
        
        \node[latent, above=of cov, xshift=-0.5cm] (alpha)  {$\alpha$};
        \node[latent, above=of cov, xshift= 0.5cm] (length) {$l$};
        
        \edge {alpha, length}       {cov};
        \edge {mu-wb, mu_farm, cov} {SGP};
        \edge {SGP}                 {bird-deaths};
    
    \end{tikzpicture}
  }
  \caption{DAG for the HPAI case study model implemented as a continuous-time individual level model with \gemlib}
  \label{fig:hpai-dag}
\end{figure}
We demonstrate the \code{ContinuousTimeStateTransitionModel} by implementing a model for Highly Pathogenic Avian Influenza (HPAI). Common models typically use a discrete-time, individual-based, spatially explicit 5-state model \cite{Davis2025Hpai}. Notably, at any given time, an individual poultry premise is categorized as susceptible to infection $S$, exposed to infection $E$, infected and able to transmit infection $I$, notified as infected $N$, and removed by culling $R$. The infectious pressure is a combination from other nearby infected poultry farms (function of covariates about each farm) and an underlying background infectious component that captures remaining pressure. The background component is used to attribute infections to wild bird spill over. 

We make two notable changes to the model described above for this case study. Firstly, we condense this to 4-state SEIR model in an effort to reduce the amount of code shown. Secondly, we introduce a spatial Gaussian Process (GP) for the background infectious pressure component of the model. The GP provides a flexible model that allows us to incorporate ecological data about wild bird HPAI prevalence\footnote{data set contains records of wild birth deaths and HPAI tests along with the result}. This is pooled with epidemiological case data in an effort to attribute infections to \emph{either} wild birds-to-farm or farm-to-farm. For each farm $j$ at time $t$, the exposure rate ($S \rightarrow E$) is assumed to be proportional to the proximity of farm $j$ to nearby infected farms $i$. The infection and removal rates  ($E \rightarrow I$, $I \rightarrow R$ respectively) are assumed to be constant. The set of transition rate functions are thus:
\begin{align}
    \lambda_{j}^{SE}(t) &=  e^{\mu_{f|j} + \textcolor{blue}{s_j}} +  \beta \sum_{i \in I_t} e^{\frac{||x_i - x_j||^2}{\phi^2}} \\
    \lambda_j^{EI}(t) &= \eta \\
    \lambda_j^{IR}(t) &= \gamma
\end{align}
Where $\rho(j,i; l, \phi) $ is the squared exponential spatial kernel with spatial decay rate $\phi$. The background term $e^{\textcolor{blue}{s_j}}$ (sometimes referred to as the spark term) is the spatial GP constructed with the Matern 3,2 covariance kernel. 
\begin{align}
    SGP &\sim MVN \left( \vec{0}, \Sigma^2 \right) \\
    \Sigma_{ij}^2 &= \mbox{Matern32}(\alpha, l) \\
    y_{i;i\in wb} &\sim \mbox{Binom}(10, \mbox{logit}^{-1}(\mu_{w.b.|i}+ \textcolor{blue}{s_i}))
\end{align}

Following a similar pattern as the previous case study, we define the incidence matrix and initial population. The population size is set to 320 farms and 80 locations for wild bird observations. Using the built-in continuous-time state transition model class, we specify the model within TensorFlow Probability's \code{tfp.Root.JointDistributionCoroutineAutoBatched}.The model DAG is shown in Figure \ref{fig:hpai-dag} and closely matches the model specification code below. The GP is defined over \emph{both} farm and wild bird locations, hence linking the ecological and epidemiological datasets.
\begin{minted}[highlightlines={16-21,68-73}]{python}
def hpai_seir_spec():
    # Gaussian process
    gp_amplitude = yield tfd.Gamma(
        concentration=np.float64(2.),rate=np.float64(1.), name="gp_amplitude"
    )
    gp_length_scale = yield tfd.Gamma(
        concentration=np.float64(16.),
        rate=np.float64(2.),
        name="gp_length_scale",
    )
    matern = tfp.math.psd_kernels.MaternThreeHalves(
            amplitude=gp_amplitude,
            length_scale=gp_length_scale,
            name="cov_kernel",
        )
    risk_surface = yield tfd.GaussianProcess(
        kernel=matern,
        index_points = jnp.concat(
            [farm_coords, wild_bird_coords]
            ),
        name="risk_surface",
    )

    # Split risk surface by species
    farm_risk_surface = jnp.asarray(risk_surface[0: NUM_FARMS])
    wb_risk_surface = jnp.asarray(risk_surface[NUM_FARMS:])

    # Wild birds model (generates ecological data)
    wb_mean = yield tfd.Normal(
        loc=np.float64(0.), scale=np.float64(2.1), name="wb_mean"
    )
    wild_bird_deaths = yield tfd.Binomial(
        total_count=10,
        logits=wb_mean + wb_risk_surface,
        name="wild_bird_deaths",
    )

    # Farms model (generates epidemiological data)
    spatial_baseline = yield tfd.Gamma(
        concentration=np.float64(1.0),
        rate=np.float64(2.0),
        name="spatial_baseline",
    )
    spatial_decay_rate = yield tfd.Gamma(
        concentration=np.float64(2.0),
        rate=np.float64(4.0),
        name="spatial_decay_rate",
    )
    farm_wb_exposure = yield tfd.Normal(
        loc=np.float64(0.), scale=np.float64(2.1), name="farm_wb_exposure"
    )

    def si_rate_fn(t, state):
        farm_farm = spatial_baseline * jnp.matvec(
            spatial_decay(pairwise_dist, spatial_decay_rate), state[:, 2]
        )
        wb_farm = jnp.exp(farm_wb_exposure + farm_risk_surface)
        return farm_farm + wb_farm

    ei_rate = yield tfd.Exponential(rate=np.float64(0.25), name="ei_rate")
    def ei_rate_fn(t, state):
        return ei_rate

    ir_rate = yield tfd.Exponential(rate=np.float64(0.25), name="ir_rate")
    def ir_rate_fn(t, state):
        return ir_rate

    farm_cases = yield ContinuousTimeStateTransitionModel(
        incidence_matrix=incidence_matrix,
        initial_state=initial_state,
        transition_rate_fn=[si_rate_fn, ei_rate_fn, ir_rate_fn],
        num_steps=600,
        name = "hpai_seir"
    )
\end{minted}

\begin{figure}[ht]
    \centering
    \includegraphics[width=0.75\linewidth]{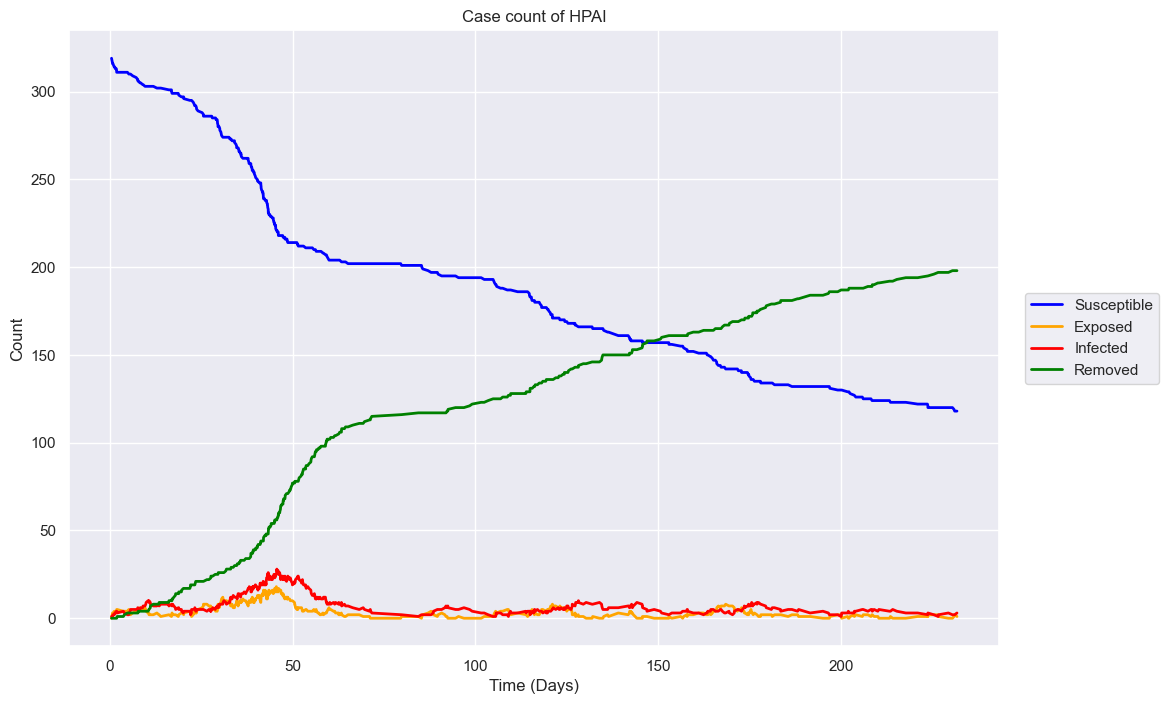}
    \caption{Simulated case count time series of the continuous-time SEIR model for HPAI. The simulation lasts a period of approximately 9 months with an initial wave in the second month followed by smaller outbreaks later in the process. }
    \label{fig:hpai_case_study_epi_curves}
\end{figure}

\begin{figure}[ht]
    \centering
    \includegraphics[width=0.95\linewidth]{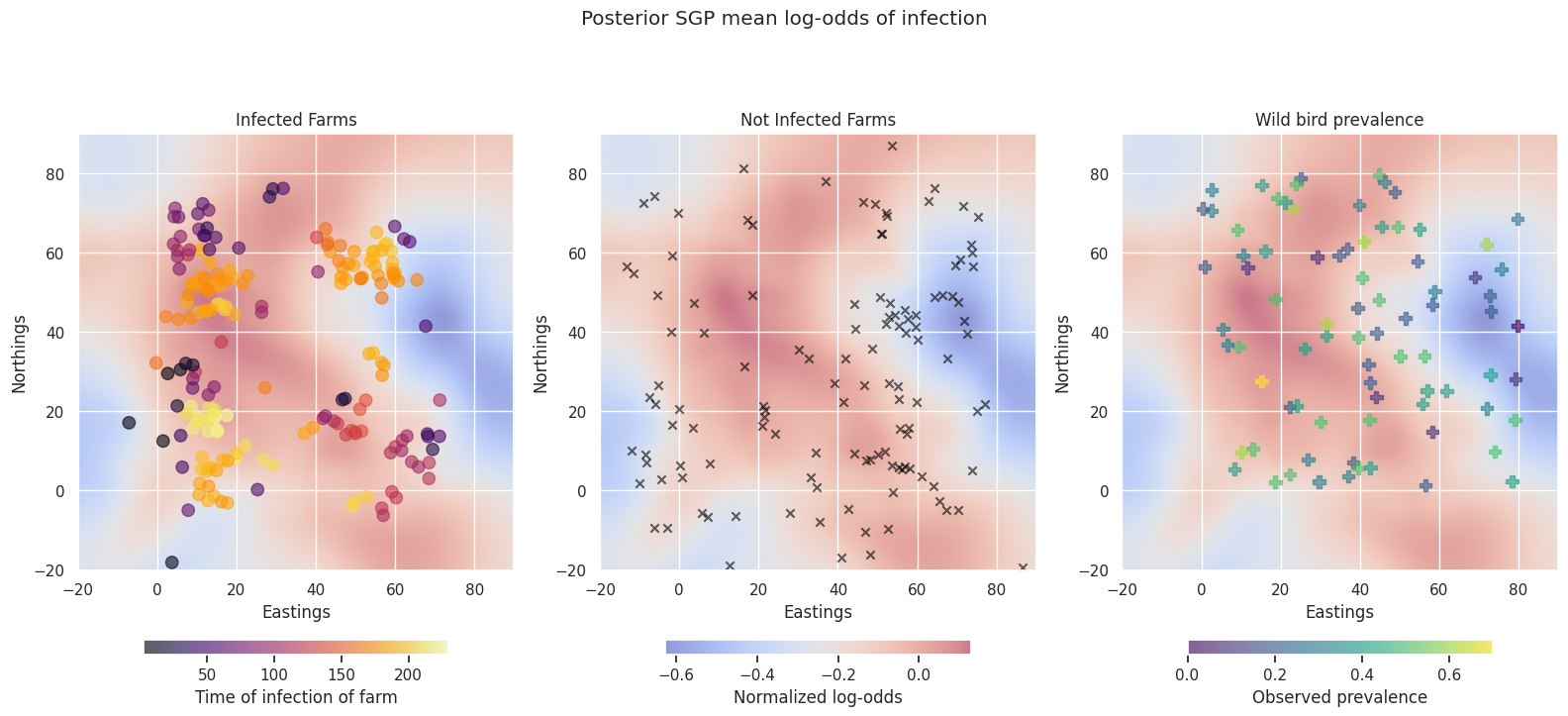}
    \caption{Plot of the spatial risk surface for simulated HPAI epidemic. The underlying risk surface shows areas of higher and lower risk based on wild bird HPAI prevalence observed being higher or lower than the average prevalence. Farms are separated across the left two plots in order to improve visibility of where infections happen. The shading of the infected farms corresponds to the time of infection. The first wave of infection is shown in the darker, purple while the later waves are lighter in colour in orange and yellow. Since the early infectious had recovered by day 150, later infections on the left side of the plot are likely due to spillover from the wild birds.}
    \label{fig:hpai_case_study_risk}
\end{figure}

We can now simulate from the model by simply passing in set parameter values for each \code{yield} statement and using the \code{sample_distributions} method of the constructed joint model. The resulting time series of counts are shown in Figure \ref{fig:hpai_case_study_epi_curves} and created by post-processing the resulting \code{EventList} produced by the continuous-time model class. Furthermore, we can plot the underlying spatial risk surface to highlight hot spots for interactions between farms and wild birds as shown in Figure \ref{fig:hpai_case_study_risk}.
\begin{minted}[linenos]{python}
# build the model with TensorFlow Probability
hpai_seir_model = tfd.JointDistributionCoroutineAutoBatched(
    hpai_seir_spec, use_vectorized_map=False
)

simulation_dists, simulation_result = hpai_seir_model.sample_distributions(
    gp_amplitude = jnp.asarray(0.25),
    gp_length_scale = jnp.array(40.),
    wb_mean=jnp.array(-0.5),
    spatial_baseline = jnp.array(1.1),
    spatial_decay_rate = jnp.array(5),
    farm_wb_exposure=jnp.array(-7.0),
    ei_rate = jnp.array(0.3),
    ir_rate = jnp.array(0.14),
    seed=jr.key(3),
)
\end{minted}

 
\begin{figure}[ht]
    \centering
    \includegraphics[width=0.70\linewidth]{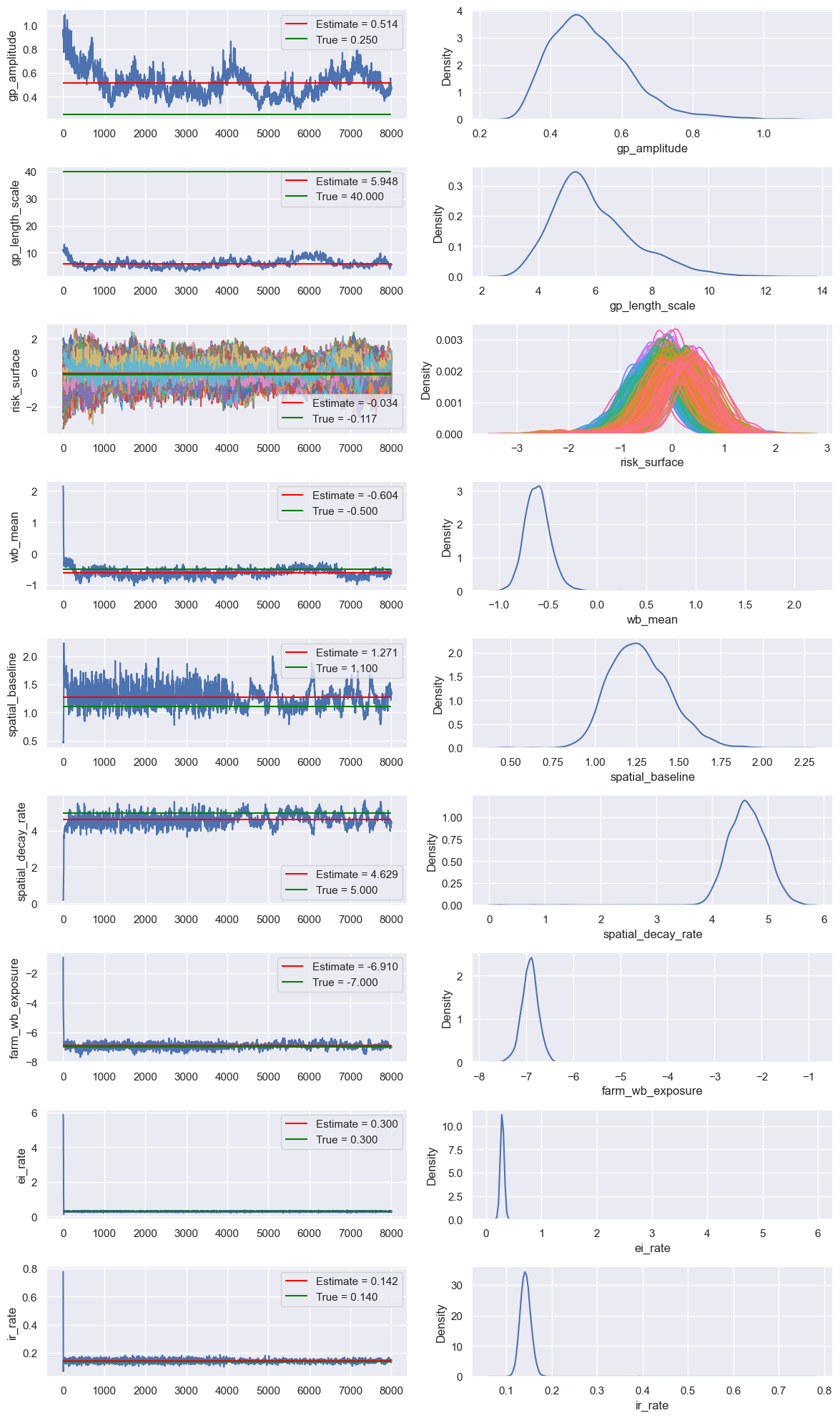}
    \caption{Trace plots for Hamilton Monte Carlo algorithm used to fit a fully observed SEIR continuous-time model.}
    \label{fig:hpai_case_study_mcmc}
\end{figure}
 
Notably, we can use the \code{hpai_sir_model} in a Bayesian inference scheme given observed data to estimate model parameters. This is equivalent to traversing "up" the DAG in Figure \ref{fig:hpai-dag}. Parameter inference was performed using an adaptive Hamiltonian Monte Carlo algorithm (trace plots show in Figure \ref{fig:hpai_case_study_mcmc}). The full code is available in the accompanying Jupyter Notebook. 

\section{Discussion}
In this paper, we introduced \gemlib, a Python library that provides a probabilistic programming interface for epidemic models. The package allows concise and structured implementation of compartmental models, while supporting seamless integration into Bayesian inference workflows. A single model specification can be used for both simulation and calibration, eliminating discrepancies between code used for forward simulation and that used for parameter estimation. Simulation is performed using GPU-optimized variants of Gillespie’s algorithm, and inference is supported through a variety of state-of-the-art methods, including approximate Bayesian computation and data augmentation MCMC. By abstracting the construction of the Markov transition kernel, \gemlib; integrates naturally with high-performance computing frameworks such as TensorFlow and JAX, allowing users to focus on scientific questions of model selection and evaluation while still benefiting from highly optimized back end. 

Importantly, \gemlib~directly addresses all concerns mentioned in Section \ref{sec:intro}. The concise model representation gives the analyst a simple way to express a wide variety of epidemic models, and effortlessly switch between continuous- or discrete-time stochastic, or deterministic processes. This models construction also outlines a clear way for \textit{how} algorithms can interact with models meaning that developers can focus on designing novel algorithms and integrate them into the library seamlessly. Lastly, by building \gemlib~on top of high performance computing libraries such as JAX and Tensorflow, models are portable and performant across different hardware architectures. Accelerators such as GPUs can be used without any modifications to the codebase. 

\subsection{Comparison with other packages}

A wide range of software frameworks exist to support epidemic modelling , but their design typically couples model specification tightly with the algorithms used for simulation and inference. The prevailing software pattern is that a specific model (such as the SIR model) is provided by the library while the user is responsible for specifying the the dynamics of the model. This approach is not conducive to an iterative model development. Common modifications such as adding a new state or otherwise modifying the model structure often requires shifting to a different module within the same library, or, in some cases, adopting an entirely new library. These type of changes require re-testing and re-validating the model and associated algorithms, all of which slows down the process of delivering timely results in a potentially emergency scenario. The principle extends to adapting models for new hypotheses or new data streams where modifications require extending the existing framework by modifying source code. 

Recent projects in both the R and Python ecosystems illustrate these limitations. Within R, the epiverse-TRACE project provides a collection of interoperable packages aimed at providing an end-to-end pipeline for analysing epidemic data \cite{eggo2024epidemics}. Model implementations are scattered throughout the ecosystem. \code{epidemics} provides simulation capabilities for predefined compartmental models and allows for encoding of intervention measures. \code{epichains} also provides simulation capabilities of some models through the implementation of branching process \cite{azam2024epichains}. Model fitting is allocated the \code{monty-odin} where models are specified as ordinary differential equations (ODEs) and are calibrated using numerical solvers or using particle filters \cite{fitzjohn2024odin}. This ecosystems showcases the disjoint nature of epidemic modelling software where similar models are found in different parts of the library. The rigidity of this approach is highlighted in the documentation for the \code{epidemics} sub-package. Developers provide instructions for modifying the underlying code with common changes needed for creating better, more representative infectious disease models. We instead advocate for a unified representation of the mathematical structure that can be reused across the entire ecosystem. Researchers can then quickly modify models at a high-level interaction with the library and minimize the number of places where changes need to trickle down through.

Outside of epiverse-TRACE, \code{InFER} is another R package which offers both simulation and inference algorithms with a GPU-compatible implementation. However, the algorithms were designed for a specific use case outlined in \cite{probertEtAl2018}, thus limiting their generalizability \cite{jewell2015infer}. Reusing the codebase in a different setting would involve rewriting portions of a C++ codebase which is a significant barrier in the epidemic modelling community.
 
The Python landscape is more diverse but similarly constrained. \code{Eir} focuses on stochastic simulations of discrete time epidemics with a heavy emphasis on spatial models \cite{JacobEIR2021}. The library adopts a philosophy of providing users with a wide array of predefined compartmental models, each with its own specific interface. The package does not provide inference capabilities. \code{Pyfectious} \cite{Mehrjou2023Pyfectious} represents infectious units (ranging from individual people all the way up to cities) as agents and simulates their interactions, yet offers no tools for parameter estimation. The library is heavily influenced by Covid-19 models and therefore imposes some limits on adaptability to other data types such as veterinary diseases. \code{EpiLearn} is constructed on top of the Python-based machine learning framework PyTorch which offers similar computational advantages as JAX. They focus on using epidemic models as part of larger processes such as forecasting and source detection \cite{liu2024epilearn}. However the primary focus is on case count data with deep learning models for prediction of time series or graph convolutional networks for modelling disease spreading. This means that the underlying compartmental models are merely a (spatio-temporal) label of the features space of the deep learning models. \code{Epiabm} re-implements the CovidSIM model which is age structured and spatial heterogenous \cite{gallagher2024epiabm} model. They adopt a modular structure to allow for simulation of a variety of interventions. The package does not provide any inference capabilities. The recently released \code{epydemix} package provides the nearest feature set to \gemlib\; since models can be both simulated and calibrated to data via approximate Bayesian computation (ABC) \cite{Gozzi2025Epydemix}. The package takes an object-oriented approach to model building where users change different properties of model that are used within general purpose simulation and inference algorithms. The library is built on top of the popular numerical computational Python library \code{numpy} and therefore lacks the high-performance capabilities we are prioritizing by using machine learning frameworks such as Tensorflow Probability and JAX. Furthermore, inference is limited to ABC methods. 

\gemlib\; was designed from the outset to decouple the model from its computational routines. A single model representation supports both simulation and inference, whether likelihood-based or simulation-based, without rewriting core code. This design ensures that researchers retain full control over model structure, can rapidly prototype novel epidemiological hypotheses, and can scale their analyses using GPU-enabled backends. The result is a framework that prioritizes flexibility, reproducibility, and extensibility—qualities essential for modelling infectious disease dynamics in real time during emerging outbreaks.

\subsection{Limitations}
At present, \gemlib; has several limitations. It does not support non-Markovian processes, restricting applications to systems with memoryless transition dynamics (assumed exponential wait times between events). The current framework assumes closed populations, with no births or immigration events, and represents deaths only as transitions to an absorbing compartment. It also assumes static (meta)populations, preventing the modelling of migration between subpopulations over time. These restrictions stem from the current design of the transition kernel abstraction, and addressing them will require extending the library to allow for composable kernels that can be chained together to perform each task.

In the implementation of the stochastic model classes in \gemlib~we made a deliberate trade-off between software engineering principles and computational performance. Notably, there is a disagreement between the method signatures of the \code{ContinuousTimeStateTransitonModel} and \code{DiscreteTimeStateTransitonModel} classes. Although Python is inherently an object-orient language, \gemlib~adopts a largely functional programming style in order to promote modularity, clarity, and performance of models. In principle, both model classes could inherit from a base class \code{StateTransitionModel} to promote code reuse and maintainability. However, such an implementation would contradict the SOLID design principles laid out by \cite{Martin2017solidsoftware}, particularly the Liskov Substitution Principle, as the data structures returned by the continuous- and discrete-time classes are not substitutable (in fact, the program would no longer run if the outputs were to be substituted for one another). This incompatibility arises from fundamental differences in how each model type encodes state changes. We therefore consciously diverge from strict object-oriented design in favour of computational performance. Materializing and storing the full epidemic state, especially for large and potentially sparse systems such as individual level models, would impose substantial memory overhead. Consequently, the present design prioritizes efficiency over formal software abstraction.

\subsection{Future directions}
Future development of \gemlib \; will aim to expand on the types of models available to users and add built-in visualization tools. High priorities include adding support for non-Markovian processes, relaxing the closed-population assumption to allow births and migration, and incorporating common visualization and diagnostic tools. A more extensive treatment of the composable inference algorithms used in \gemlib; will be provided in a future publication. As the library evolves, we anticipate that its modular architecture, coupled with deep integration into high-performance computing ecosystems, will make it a valuable platform for researchers developing novel models and inference methods in epidemic modelling at scale.

\appendix
\section{Continuous time models} \label{supp:cts-time-model}
Continuous-time, inhomogeneous Poisson processes offer a robust mathematical framework for modelling the stochastic nature of epidemic events occurring at irregular intervals (rendering them particularly well-suited for individual-level models). These models naturally accommodate fine-grained, time-varying covariates and heterogeneous transition rates shaped by contextual factors such as contact networks \cite{Bridgen2024-ao} or individual-level attributes \cite{Dangerfield2009}.

A defining feature of continuous-time models is that events occur asynchronously: within any time interval $[t, t+\delta t)$, at most one event can take place. This reflects the reality of epidemic transitions which are inherently stochastic and mutually exclusive at the individual level. In this framework, the exact time since the last event, $\delta t$, is explicitly tracked. We assume that, conditional on the current population state $x_t$, individuals are independent within each time interval $[t, t+\delta t)$ thus making this a Markov model \cite{keeling2011modeling}. It then follows that for each individual, the set of possible (Markovian) transitions are set by the evaluation of the transition functions in $\Lambda_\mathcal{Z}$. 

The stochastic dynamics can be simulated via Gillespie's algorithm \cite{gillespie1977} which is widely used for simulation of continuous-time Markov processes in population dynamics. Let $S$ be the number of Markov jumps to simulate\footnote{Typically, the Gillespie algorithm continues until there are no possible transitions. This is done with a \texttt{while} loop that checks if $\sum \Lambda_\mathcal{Z} > 0$. In an effort to unify the continuous and discrete time algorithms, we change the continuous time algorithm to run for a fixed number of jumps.}. The details are listed in Algorithm \ref{alg:cts-gillespie}.

\begin{algorithm}[ht]
    \caption{The Gillespie algorithm for simulating continuous-time epidemic models} \label{alg:cts-gillespie}
    
    \Initialize: $\mathbf{x}_0 \in \mathbb{N}^{|N| \times |\mathcal{X}|}, \mathbf{B}$
    
    \For{$s=1,\dots,S$}{
    Compute $ \lambda^{\stx{q}{r}}(t_s, \mathbf{x}_s)$ $\forall \tx{q}{r}\in \mathcal{Z}_{SIR} $ and $\Lambda = \sum_{\tx{q}{r}\in \mathcal{Z}_{SIR}} \lambda^{\stx{q}{r}}(t_s, \mathbf{x}_s)$\;
    Sample time to next event $\delta t \sim \mbox{Exp} (\Lambda)$ \;
    Sample event $k \sim \mbox{Discrete}\left( \frac{\lambda^{\stx{q}{r}}}{\Lambda} \right)$ \; 
    Update $t_{s+1}$ and $\mathbf{x}_{s+1}$ accordingly
    }
\end{algorithm}
The state-update is performed by doing a row addition according to the equation below.
\begin{equation}
    \mathbf{x}^{N \bmod k, \cdot}_{s+1} = \mathbf{x}^{N\bmod k, \cdot}_{s} + (\mathbf{B}^{\cdot, N \text{ rem } k})^T
\end{equation}
This provides a simple and fast computation since the coordinates of the row being updates are explicit.

This simulation-based perspective leads directly to the likelihood function for continuous-time stochastic epidemic models. Each event is a Markov jump, defined by two random components:
\begin{enumerate}
    \item an exponentially distributed waiting time
    \item a discrete choice of transition and individual which is encoded as a categorical random variable
\end{enumerate}

Thus, the probability of observing a transition of type $\tx{q}{r}$ involving individual $j$ at time $t$, given the population state $\mathbf{x}_t)$, is:
\[
\mathbb{P}(x^{\stx{q}{r}}_j = 1 \mid t, \mathbf{x}_t)) = \lambda^{\stx{q}{r}}_j(t, \mathbf{x}_t)) \cdot \exp\left\{-\delta t \sum_{q,r,j} \lambda^{\stx{q}{r}}_j(t, \mathbf{x}_t))\right\}
\]

The likelihood of observing a full sequence of such transitions is then:
\begin{align}\label{eq:cts_likelihood}
    \mathcal{L}(\theta ; [X]_t) &=  \prod_t \frac{\lambda^{\stx{q}{r}}_j(t)}{\sum_{q,r,j} \lambda^{\stx{q}{r}}_j(t)} \cdot \sum_{q,r,j} \lambda^{\stx{q}{r}}_j(t) \exp \{-\delta t \sum_{q,r,j} \lambda^{\stx{q}{r}}_j(t) \} \\
    &= \prod_t \lambda^{\stx{q}{r}}_j(t) \cdot \exp \{-\delta t \sum_{q,r,j} \lambda^{\stx{q}{r}}_j(t) \}
\end{align}

This expression forms the basis for likelihood-based inference, linking the observed data to model parameters through a DAG similar to Figure \ref{fig:sir-dag}.

\section{Discrete time models}\label{supp:disc-time-model}
In practice, epidemiological concerns are centred around the number of cases, hospitalizations, or deaths per day \cite{Bootone2021epicasesSW} and as such, there is a natural discretization of time. This is also beneficial from a computational perspective since bundling events can be used to accelerate both simulation and likelihood evaluation. 

The continuous-time model can be approximated by discretising time into intervals of fixed length $\delta t$, giving rise to the \emph{chain-multinomial model}. his approximation is useful when event data is observed at fixed intervals (e.g., daily or weekly) or when computational constraints prevent use of the preferred continuous-time paradigm \cite{keeling2011modeling}. 

In this discrete-time approximation, the survival probability (i.e., the probability of \emph{not} undergoing a $\tx{q}{r}$ transition) for a unit $j$ is:
$$\mathbb{P}(x^{\stx{q}{r}}_j = 0 \mid t, \mathbf{x}_t)) = 1 - e^{-\delta t \cdot \lambda^{\stx{q}{r}}_j(t, \mathbf{x}_t))}$$

We assume that within an interval $[t, t+\delta t)$, all individuals are conditionally independent given the population state $\mathbf{x}_t$ at time $t$.  For each population unit, we define the transition rate matrix $Q$, where the off-diagonal $q,r$th elements are the $\lambda^{qr}(t, \mathbf{x}_t))$, and diagonal elements are given by the negative of the row-wise off-diagonal elements.  The $j^\text{th}$ row of $Q$ is 
\begin{equation}
    Q_j(t, x_t) = \left[ \begin{array}{llll}
           \ldots 0 \ldots & -\lambda^{\stx{q}{r}}(t, \mathbf{x}_t) & \lambda^{\stx{q}{r}}(t, \mathbf{x}_t) & \ldots 0 \ldots \\
        \end{array} \right]
\end{equation}
from which one can obtain the right-stochastic matrix $P(t) = e^{Q(t) \delta t}$. However, since matrix exponentials are slow to compute, we make a further simplifying assumption that individuals may only undergo a maximum of \emph{one} transition per timestep.  This allows us write the stochastic matrix as
\begin{equation}
P_j(t, x_t) = \left[ \begin{array}{llll}
           \ldots 0 \ldots & e^{-\lambda^{\stx{q}{r}}(t, x_t) \delta t} & 1 -  e^{\lambda^{\stx{q}{r}}(t, x_t) \delta t} & \ldots 0 \ldots \\
        \end{array} \right]
\end{equation}

The data-generating process then uses multinomial distributions applied row-wise to $P$. For each source compartment $q$, we draw the number of $[qr]$ transitions from a multinomial random variable with $x^q_t$ trials and probability vector $\mathbf{p}^{q}(t, \mathbf{x}_t)$, the $q^{\text{th}}$ row of $P$:
\begin{equation}
    \mathbf{z}^{q\cdot}_t \sim \mathrm{Multinomial}(x^{q}_t, \mathbf{p}^{q}(t, \mathbf{x}_t))
\end{equation}
The updated state vector $\mathbf{x}_{t+\delta t}$ at time $t + \delta t$ is then:
\begin{equation}
    \mathbf{x}_{t+\delta t} = \sum_q \mathbf{z}^{q\cdot}_t
\end{equation}

This discrete-time formulation corresponds to the classical chain-binomial algorithm introduced by \textcite{becker81}.  However, the multinomial algorithm allows for more general state transition models to be specified including branches and cycles as required by the specific application. The state update is computed according to the matrix product of the incidence matrix $\mathbf{B}$ and the state change $\mathbf{z}^{q\cdot}_t$ as described in \cite{Black2019Importance}. The algorithm for simulating $S$-many Markov jumps of the discrete-time model proceeds as follows: 

\begin{algorithm}[ht]
    \caption{The chain-multinomial algorithm for simulation discrete-time epidemic models} \label{alg:chain-multinomial}
    
    \Initialize: $\mathbf{x}_0 \in \mathbb{N}^{|N| \times |\mathcal{X}|}, \mathbf{B}$
    
    \For{$s=1,\dots,S$}{
    Compute $P(\mathbf{x}_s; \lambda^{\stx{q}{r}}(s, x_s))$ \;
    \ForEach{$q \in \mathcal{X}$}{
      Sample $\underline{z}_{s+1}^{q\cdot} \sim \mbox{Multinomial}(x^{q}_s, \mathbf{p}^{\mathbf{q}}(t_s, \mathbf{x}_s))$
    }
    Update state $\mathbf{x}_{s+1}$ with $ z_{s+1}$ \; 
    }
\end{algorithm}

The multinomial distribution gives the probability that all units in each compartment of $\mathcal{X}$ transition according to probabilities derived from the rows of $P$. For each transition (i.e. the number of transition events $k^{\tx{q}{r}} \sim \mbox{Multinom}(\mathcal{X}_q, p_{\stx{q}{r}}(t)) $). The likelihood function for the full sequence of observations is the product over all time steps and transitions, assuming the convention that $0^0 = 1$.
\begin{equation}
    \mathcal{L}(\theta ; [X]_t) = \prod_{t} \prod_{\tx{q}{r}\in \mathcal{Z}} \frac{n_t!}{z^{\tx{q}{r}}_t!} \times p_{\stx{q}{r}}(t, \mathbf{x}_t)^{{z^{\tx{q}{r}}_t}}
\end{equation}

This formulation captures the probabilistic structure of epidemic processes. By representing transitions as competing exponential hazards or aggregated multinomial trials, the model supports parameter inference using maximum likelihood or Bayesian methods. The factorization into survival probabilities and multinomial weights enhances computational efficiency while remaining faithful to the underlying stochastic dynamics.

\begin{figure}[h!]
    \centering 
        \begin{tikzpicture}
            \filldraw[fill=green!20!white, draw=black] (0,0) rectangle node{\large S} (2,2) ;
            
            \draw[thick,->] (2.5,1) -- node[anchor = south]{$\lambda^{SI}(t, X_t)$} (4.5, 1) ;
            
            \filldraw[fill=lancasterRed!20!white, draw=black] (5,0) rectangle node{\large I}(7,2);
    
            \draw[thick,->] (7.5,1) -- node[anchor = south]{$\lambda^{IR}(t, X_t)$} (9.5, 1) ;
            
            \filldraw[fill=lancasterLightBlue!20!white, draw=black] (10,0) rectangle node{\large R} (12,2);
    
        \end{tikzpicture}
    \caption{Susceptible-Infected-Removed compartmental model, lines represent the transition events, functions above the line indicate the transmission rates}
    \label{fig:SIR}
\end{figure}
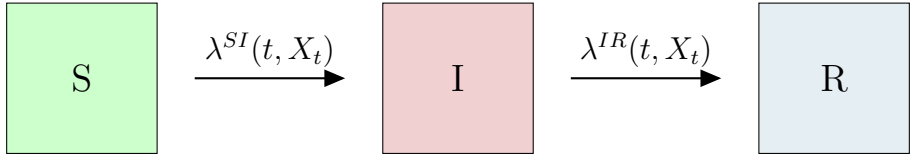

\section{SIR model}
The canonical model is the Susceptible-Infected-Removed (SIR) model first introduced by \cite{kerMcK1927}. In this model epidemic units are categorized within three compartments. Each unit belongs to \textit{only} one of the disease states at any given point in time. They may be \textbf{S}usceptible (unexposed to the pathogen), \textbf{I}nfected (currently colonized by the pathogen), or \textbf{R}emoved (if they have recovered from the infection or otherwise not infectious anymore) \cite{keeling2011modeling}. We track the epidemic by keeping count of the number of units in each compartment using a state-vector $\mathcal{X}_t = (S_t, I_t, R_t)$ which represents the state of the system at time $t$. An infection event indicates a unit moved from $S \rightarrow I$ while a removal event sees a unit move from $I \rightarrow R$ and therefore we can define the transitions as $\mathcal{Z} = \{ \tx{S}{I}, \tx{I}{R} \}$. This is conveniently summarized with a flow diagram as shown in Figure \ref{fig:SIR}. 

For this section, we will focus on the frequency depended model where the transition rates are proportional to the number of possible interactions between epidemic units. The rates are as follows: 
\begin{align}
    \lambda^{\stx{S}{I}}(t, \mathbf{x}_t) &= \beta x_t^S x_t^I \\
    \lambda^{\stx{I}{R}}(t, \mathbf{x}_t) &= \gamma
\end{align}
Where the model parameters are $(\beta, \gamma)$. Note that the transition rate functions are identical for both the continuous and discrete time models. The remainder of this section outlines the simulation and probability evaluation of this model in each time domain.

\subsubsection{Continuous time}
Simulation is performed using an instance of Gillespie's algorithm shown in \ref{alg:cts-gillespie}.
\begin{algorithm}[ht]
    \caption{The Gillespie algorithm for simulating continuous-time, population level SIR model} \label{alg:cts-gillespie_sir}
    
    \Initialize: $\mathbf{x}_0 \in \mathbb{N}^{3}, B$
    
    \For{$s=1,\dots,S$}{
    Compute $\lambda^{\stx{S}{I}}(t_s, \mathbf{x}_s), \lambda^{\stx{I}{R}}(t_s, \mathbf{x}_s)$ and $\Lambda = \lambda^{\stx{S}{I}}(t_s, \mathbf{x}_s) + \lambda^{\stx{I}{R}}(t_s, \mathbf{x}_s)$\;
    Sample time to next event $\delta t \sim \mbox{Exp} (\Lambda)$ \;
    Sample event $k \sim \mbox{Discrete}\left( \frac{\lambda^{\stx{S}{I}}}{\Lambda}, \frac{\lambda^{\stx{I}{R}}}{\Lambda} \right)$ \; 
    Update $t_{s+1} = t_s + \delta t$ and $\mathbf{x}_{s+1} = \mathbf{x}_{s} + B_{\cdot K}^T$ 
    }
\end{algorithm}
The total event rate $\Lambda$ depends only on the two transition rates and the state update is performed by adding the $K^{\text{th}}$ column of the incidence matrix.

\subsection*{Probability}
The likelihood function for the model is derived by taking the product of the two random variables seen in Algorithm \ref{alg:cts-gillespie_sir}. The first is the waiting time to the next event which is exponentially distributed. The second is the type of event that occurred which is a categorical random variable. The joint-probability of all events $H$ can be evaluated by taking the product over the time stamps of each event. 
\begin{align}
    L(X|H) &= \prod_{s=0}^{max - 1} \frac{\lambda(t_s)}{\lambda^{SI}(t_s)+\lambda^{IR}(t_s)} \cdot \left( \lambda^{SI}(t_s)+\lambda^{IR}(t_s) \right) e^{ -\left(\lambda^{SI}(t_s)+\lambda^{IR}(t_s)\right) \delta t_s} \\
    &= \prod_{s=0}^{max - 1} \lambda(t_s) e^{ -\left(\lambda^{SI}(t_s)+\lambda^{IR}(t_s)\right) \delta t_s} 
\end{align}

\subsection{Discrete time}
Similarly with the discrete time variants, we take an instance of the chain multinomial algorithm \ref{alg:chain-multinomial}. There are two transitions events since $\mathcal{Z}$ has elements and so there are two underlying binomial experiments. To simulate this process, we sample the number of $\tx{S}{I}$ and $\tx{I}{R}$ events respectively as shown in Algorithm \ref{alg:chain-multinomial-SIR}. For simplicity, we set $\delta t = 1$ unit of time. 
\begin{algorithm}[ht]
    \caption{The chain-binomial algorithm for simulating discrete-time, population level SIR model} \label{alg:chain-multinomial-SIR}
    
    \Initialize: $\mathbf{x}_0 \in \mathbb{N}^{3}, B$
    
    \For{$s=1,\dots,S$}{
    Compute $p^{SI}_{s}=1-e^{-\lambda^{\stx{S}{I}}}, p^{IR}_{s}=1-e^{-\lambda^{\stx{I}{R}}(s, \mathbf{x}_s)}$ \;
    Sample $z_{s+1}^{SI} \sim \mbox{Binomial}(x^{S}_s, p^{SI}_{s})$, $z_{s+1}^{IR} \sim \mbox{Binomial}(x^{I}_s,p^{IR}_{s}) $ \;
    Update state $\mathbf{x}_{s+1} = \mathbf{x}_{s} + (z_{s+1}^{SI}, z_{s+1}^{IR}) \cdot B^T$ \;
    }
\end{algorithm}
\subsection*{Probability}
The joint probability of events in the process is the product of the two binomial distributions. 
\begin{align}
L(X|H) &= \prod_{s = 0}^{max - 1} \binom{x^{S}_s}{z_{s+1}^{SI}}  \left(p^{SI}_{t_s} \right)^{z_{s+1}^{SI}} \left(1-p^{SI}_{t_s} \right)^{x^{S}_s-z_{s+1}^{SI}} \cdot  \binom{x^{I}_s}{z_{s+1}^{IR}} \left(p^{IR}_{t_s} \right)^{z_{s+1}^{IR}} \left(1-p^{IR}_{t_s} \right)^{x^{I}_s - z_{s+1}^{IR}} \\
&\propto \prod_{s = 0}^{max - 1}\left(p^{SI}_{t_s} \right)^{z_{s+1}^{SI}} \left(1-p^{SI}_{t_s} \right)^{x^{S}_s-z_{s+1}^{SI}} \cdot \left(p^{IR}_{t_s} \right)^{z_{s+1}^{IR}} \left(1-p^{IR}_{t_s} \right)^{x^{I}_s - z_{s+1}^{IR}}
\end{align}

\nocite{*}
\printbibliography

\end{document}